\documentclass[acmsmall,screen,nonacm]{acmart}

\usepackage{amsmath,amsfonts}
\usepackage{algorithm}
\usepackage{algpseudocode}
\usepackage{graphicx}
\usepackage{textcomp}
\usepackage{listings}
\usepackage{xcolor}
\usepackage[english]{babel}
\usepackage{color,colortbl}
\usepackage{xspace}
\usepackage{multirow}
\newcommand{\evshade}{\cellcolor{black!5}}  
\usepackage{graphicx}
\usepackage{booktabs}
\usepackage{url}
\usepackage{xurl}
\usepackage{tipa}
\usepackage{fvextra}
\input{arxiv-code/style.tex}
\VerbatimPygments{\PYG}{\PYG}
\usepackage[T1]{fontenc}
\usepackage[utf8]{inputenc}
\usepackage{threeparttable}
\usepackage{amsmath,amsfonts}
\usepackage{pifont}
\usepackage{tabularx}
\usepackage[printonlyused]{acronym}
\usepackage[most]{tcolorbox}
\usepackage{multicol}
\usepackage{wrapfig}
\usepackage{adjustbox}
\usepackage{fontawesome5}
\usepackage{hyperref}
\usepackage{enumitem}
\hypersetup{
	pdftitle={},
	pdfauthor={},
	pdfsubject={},
	pdfkeywords={},
	bookmarksnumbered=true,
	bookmarksopen=false,
	colorlinks=true,
	pdfstartview=FitB,
	pdfpagemode=UseOutlines
}

\usepackage{tikz}
\usetikzlibrary{shapes,arrows,positioning,fit}
\usetikzlibrary{backgrounds}
\usetikzlibrary{arrows.meta, positioning, calc}
\usepackage{pgfplots}
\pgfplotsset{compat=newest}
\definecolor{vizblue}{HTML}{2A78D6}
\definecolor{vizorange}{HTML}{EB6834}
\definecolor{vizaqua}{HTML}{1BAF7A}
\definecolor{figblue}{HTML}{4F647C}   
\definecolor{figochre}{HTML}{9C7A3C}  

\definecolor{mygreen}{rgb}{0,0.6,0}
\definecolor{mygray}{rgb}{0.5,0.5,0.5}
\definecolor{mymauve}{rgb}{0.58,0,0.82}
\definecolor{pastelcyan}{rgb}{0.741, 0.867, 0.894}
\definecolor{skyice}{rgb}{0.62, 0.776, 0.953} 
\definecolor{warmlinen}{rgb}{0.949, 0.937, 0.906}
\definecolor{powderblue}{RGB}{152,193,217}
\definecolor{sageleaf}{RGB}{233, 237, 201}
\definecolor{blushcloud}{RGB}{241,231,231}
\definecolor{gainsboro}{RGB}{220,220,220}

\definecolor{mygold}{RGB}{175, 149, 0}   
\definecolor{mysilver}{RGB}{180, 180, 180} 
\definecolor{mybronze}{RGB}{173, 138, 86} 
\definecolor{sharpreviewred}{RGB}{176, 0, 0}

\newif\ifreviewcomments
\reviewcommentsfalse
\newcommand{\reviewcomment}[1]{\ifreviewcomments#1\fi}

\newcommand{\todo}[1]{\reviewcomment{\textcolor{red}{\textbf{[TODO\ifx&#1&\else: #1\fi]}}}}
\definecolor{llmstylepurple}{RGB}{128, 0, 128}
\newcommand{\llmstyle}[1]{\reviewcomment{\textcolor{llmstylepurple}{\textbf{[LLM-style\ifx&#1&\else: #1\fi]}}}}
\definecolor{termfixteal}{RGB}{0, 112, 128}
\newcommand{\termfix}[1]{\reviewcomment{\textcolor{termfixteal}{\textbf{[TERM\ifx&#1&\else: #1\fi]}}}}

\newcommand\zhiyun[1]{\reviewcomment{{\color{cyan}{--Zhiyun: #1 --}}}}

\newif\ifrevmarks
\revmarksfalse
\definecolor{revtagcolor}{RGB}{176, 0, 0}
\definecolor{revtodocolor}{RGB}{0, 110, 60}
\definecolor{cutcandidatecolor}{RGB}{138, 43, 172}

\newcommand{\eg}{\textit{e.g.,}\xspace}    

\newcommand{\tool}{\textsc{Evident}\xspace}
\newcommand{\framac}{Frama-C\xspace}
\newcommand{\eva}{Eva\xspace}

\definecolor{tkbar}{HTML}{5A7D64}    
\definecolor{tklabel}{HTML}{3F5F47}  
\definecolor{tkink}{HTML}{3A3A3C}    
\definecolor{tkfill}{HTML}{F3F8F4}   
\newtcolorbox{takeaway}[1]{enhanced, breakable, sharp corners, frame hidden,
  colback=tkfill, boxrule=0pt, borderline west={2.5pt}{0pt}{tkbar},
  left=8pt, right=6pt, top=3.5pt, bottom=4pt, boxsep=0pt,
  before skip=7pt plus 2pt, after skip=7pt plus 2pt, coltext=tkink,
  before upper={{\small\textcolor{tklabel}{\textsc{Takeaway}\enspace\textsc{#1}}}\par\nobreak\vspace{1pt}}}

\newcommand{\squishlist}{
	\begin{list}{$\bullet$} {
			\setlength{\itemsep}{0pt}
			\setlength{\parsep}{0pt}
			\setlength{\topsep}{0pt}
			\setlength{\partopsep}{0pt}
			\setlength{\leftmargin}{1.0em}
			\setlength{\labelwidth}{1em}
			\setlength{\labelsep}{0.5em}
		}
		}
		\newcommand{\squishend}{
	\end{list}
}

\newcommand{\cut}[1] {}

\cut{\zhiyun{a few things we should do before submission:
		\begin{itemize}
			\item check captilization consistency, ``figure -> Figure'', caption of a figure/table should have the first letter captilized only
			\item center the caption of all figures/tables.
			\item remove the ``period'' mark in captions.
			\item spell check / grammar check (put all sentences in grammarly)
			\item for all the functions referened in textit\{\}, we should add the brackets. For example, ``textit\{fun\} -> textit\{fun()\}''
			\item global replace of ``\ textit'' with ''\ texttt''. Also, there is no need to add the quotation marks around \texttt.
			\item check the text that references the figure and the figure itself are always co-located on the same page (or one page before/after).
		\end{itemize}
	}}

\setcopyright{none}
\acmDOI{}
\acmYear{2026}
\thanks{This is a preprint of an article submitted to a conference on planet
Earth (mostly harmless), orbiting a small star in the uncharted backwaters
of the unfashionable end of the western spiral arm of the Milky Way.\\
\copyright\ 2026 The Authors. Licensed under CC BY 4.0.}

\title{The Hitchhiker's Guide to Program Analysis, Part III: Mostly Harmless LLMs}
\author{Haonan Li}
\affiliation{\institution{University of California, Riverside}\city{Riverside}\country{USA}}
\email{hli333@ucr.edu}
\author{Tianyang Zhou}
\affiliation{\institution{University of Illinois Urbana-Champaign}\city{Urbana}\country{USA}}
\email{tz64@illinois.edu}
\author{Manu Sridharan}
\affiliation{\institution{University of California, Riverside}\city{Riverside}\country{USA}}
\email{manu@cs.ucr.edu}
\author{Hang Zhang}
\affiliation{\institution{Indiana University Bloomington}\city{Bloomington}\country{USA}}
\email{hz64@iu.edu}
\author{Zhiyun Qian}
\affiliation{\institution{University of California, Riverside}\city{Riverside}\country{USA}}
\email{zhiyunq@cs.ucr.edu}

\begin{document}

\begin{abstract}
Filtering out a static-analysis warning asserts that no feasible execution can
trigger the reported error, whereas confirming a bug requires only one that
does. Recent work uses LLMs to decide which warnings to filter out.
However, LLM-generated explanations alone cannot establish that the reported
error is impossible, and mistakenly filtering out a warning conceals a real
bug. Sound static analysis can provide this guarantee, but scalable analyses
leave behind the very warnings we seek to filter, whereas sufficiently precise
whole-program analysis is impractical. Fortunately, precisely analyzing a
small part of the program is often sufficient to show that a warning is a
false alarm. We therefore assign LLMs a different role: rather than judging
the validity of warnings, they identify relevant code and construct a harness
that supplies the context needed for precise, targeted static analysis.

We present \tool, which combines LLMs with the static analyzer \framac/\eva.
Because a generated harness may exclude inputs or paths that trigger real
bugs, \tool restricts how the harness initializes inputs and validates it
using the value flow reported by the original detector. A warning is
filtered out only when validation passes and \eva rules out the reported
error in the harness; otherwise, the warning is kept. On 200 real Android
kernel-driver warnings, including 40 confirmed bugs, \tool filters out
48.8--63.8\% of the 160 false alarms across four LLMs without missing any
confirmed bug, whereas baselines that let LLMs make the final decision miss confirmed bugs with every evaluated model. With GPT-5.5, although these baselines achieve 2.0--5.5
percentage points higher overall accuracy by filtering out more false alarms,
they miss 27.5--45.0\% of the confirmed bugs. \tool also rediscovers a
confirmed vulnerability that had previously been overlooked by both LLM-based
filtering and manual review.

\end{abstract}
\maketitle

\section{Introduction}

Static analysis tools are widely used to find bugs in large codebases such as
the Linux kernel, but they also generate many false alarms.
To reduce manual effort, recent work uses LLMs to post-process warnings, both
to confirm real bugs and to filter out false alarms~\cite{du2026reducing-fp,
llift, buglens, lekssays2025llmxcpg}.
An LLM can, in principle, confirm a bug by finding a feasible execution
that triggers the reported error; filtering out a warning, however,
requires ruling out \emph{all} such executions.
An LLM may offer a plausible explanation for a no-bug decision, but that
explanation alone cannot establish that the reported error is impossible,
and mistakenly filtering out a warning hides a real bug.

On the other hand, program analysis can provide such justification.
Sound static analysis techniques, such as abstract interpretation,
can establish that a reported error cannot occur.
However, applying them to large programs presents a familiar dilemma:
scalable analyses may be too imprecise to rule out potential errors,
leaving the very warnings we seek to filter;
analyses precise enough to rule out these false alarms may be
too expensive to run on the entire program, especially on the Linux
kernel~\cite{lawall_should_2025, chen_veld_2024}.

Fortunately, precise analysis of a small part of the program is often
sufficient to show that a warning is a false alarm.
For example, a bounds check near an array access may guarantee
that the access is safe for every possible incoming index,
regardless of how that index was computed.
The challenge is to identify the relevant code and supply the context
such analysis needs, without analyzing the rest of the program.
Rather than asking LLMs to decide whether a warning is a false alarm,
we use them to construct warning-specific harnesses\footnote{We use
\emph{harness} in the program-analysis sense, as in a test or fuzzing
harness: a program that sets up the state in which a piece of code is
analyzed. We do not mean the recent usage in which a ``harness''
orchestrates an LLM.} that supply this context, and leave the decision
to the analyzer.

However, this is not an ordinary propose-and-verify setting: the LLM's
proposal is the program the analyzer examines.
The analyzer can rule out the reported error in that harness, but it
cannot tell whether the harness preserves every execution that triggers
the error in the original program; checking that would require the
precise analysis of the original program that was impractical in the
first place.
If the harness narrows an input or drops a path that matters, the
analyzer rules out the error in the harness while the original program
still has the bug.
We therefore restrict how the harness is built and use validation
checks to detect potential problems that could hide real bugs.

We introduce \tool to implement this approach.
Given a warning and the surrounding kernel code, the LLM constructs a C
harness that provides an entry point to the relevant code and sets up
the program state needed for analysis.
Rather than letting the LLM choose concrete values for warning-relevant
numeric inputs, \tool initializes them abstractly over their full C
type ranges.
\tool then analyzes the generated harness with \framac/\eva, relying on
the analyzer rather than the LLM to determine whether the reported
error can be ruled out in the harness.

To guard against flaws in the generated harness, \tool validates it using the
value flow reported by the original detector, checking whether the harness could exclude executions triggering the reported error.
When validation fails, the LLM receives feedback to revise the harness.
\tool filters out a warning only when validation passes and \eva rules out
the reported error in the harness; otherwise, the warning is kept, including
when the analysis remains unresolved.

We evaluate \tool on 200 real Android kernel-driver warnings
reported by two static detectors, including 40 confirmed bugs
and 160 false alarms.
Across four LLMs, \tool filters out 78--102 false alarms (48.8--63.8\%)
without missing any confirmed bug, whereas the LLM-decided baselines
miss confirmed bugs with every evaluated model.
With GPT-5.5, \tool filters out 102 false alarms (63.8\%).
The corresponding LLM-decided baselines achieve 2.0--5.5 percentage
points higher overall accuracy by filtering out more false alarms,
but miss 11--18 confirmed bugs (27.5--45.0\%).
Our design-choice experiments further show that relaxing the
restrictions on LLM-generated harnesses or returning the final
decision to the LLM introduces false negatives.
\tool also rediscovers a confirmed vulnerability that had previously been
overlooked by both LLM-decided filtering and manual review.

This paper makes two contributions:
\squishlist

\item \textbf{Restricting LLM judgment in warning filtering.}
We argue that LLMs should make program analysis practical,
not judge warnings in its place.
The LLM reconstructs the context needed for analysis,
and program analysis determines whether the reported error
can be ruled out in that context.
The LLM's choices about this context are expressed in a program,
where they can be restricted and checked.

\item \textbf{Design and evaluation of \tool.}
We implement this approach in \tool, which uses LLMs to construct
warning-specific harnesses, validates them using the detector's reported
value flow, and analyzes them with \framac/\eva.
Across 200 real Android kernel-driver warnings and four LLMs, \tool
eliminates 48.8--63.8\% of the 160 false alarms without missing
any of the 40 confirmed bugs, whereas LLM-decided baselines miss
confirmed bugs with every evaluated model.

\squishend
\begin{figure}[t]
\centering
\includegraphics[width=\textwidth]{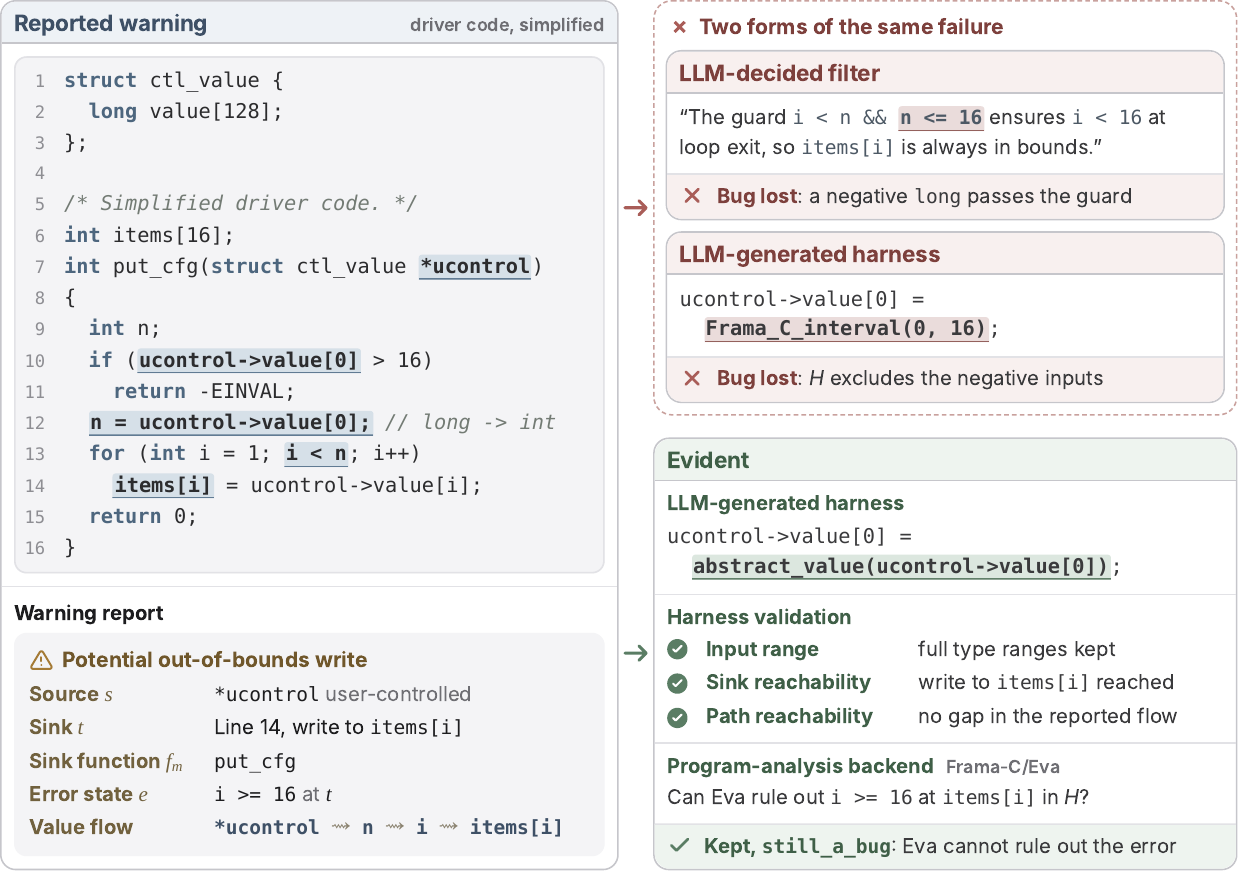}
\caption{Motivating warning and two forms of the same failure: an LLM-decided filter removes this real bug by relying on the local guard, and an LLM-generated harness encodes the same assumption by constraining the attacker-controlled input. Before analysis, \tool checks the harness properties that such errors violate.}
\label{fig:motivation-overview}
\end{figure}

\section{Motivation}
\label{sec:motivation}

We motivate \tool with a real Android kernel-driver bug that was overlooked by
both prior LLM-based filtering and manual review, but corresponds to an existing
vendor patch~\cite{audio_kernel_patch}. A negative value passed through the
user-controlled \texttt{ucontrol} structure could become a positive bound and
lead to an overflow. The example exposes the same incorrect assumption in two
forms. An LLM can use a plausible local guard to justify a no-bug judgment in
text. If the LLM is instead used to construct an analysis harness, it can encode
the same assumption into the program by constraining the attacker-controlled
input and thereby hide the bug from the backend analysis.

\subsection{A Plausible No-Bug Judgment Can Be Wrong}
\label{sec:motivation-example}

Figure~\ref{fig:motivation-overview} shows a warning that appears
easy to filter out when viewed locally. The loop bound is checked
against the size of \texttt{items}, and the write occurs immediately
after the check. An LLM-based filter, or a human reviewer reading
only the local snippet, can therefore give a convincing no-bug
rationale: the guard at Line~10 appears to bound the loop, so the
out-of-bounds warning looks like a false alarm.

The local rationale misses the type of the checked value. The value
is not an \texttt{int}, but a 64-bit \texttt{long} field in a control
object defined by ALSA (the Advanced Linux Sound Architecture), the
Linux kernel's audio subsystem. A negative \texttt{long} can pass
the guard and then be narrowed to a positive 32-bit \texttt{int}.
For example, $-(2^{32} - 17)$ becomes $17$ after the narrowing
conversion. Thus, the guard does not establish the condition needed
to make the loop safe, namely \texttt{n <= 16}, and the write to
\texttt{items[i]} can go out of bounds.

This example shows why a plausible local explanation is not enough
to filter out a warning. The visible check and the reported write
are adjacent, but the value's type and full range of possible
values are determined outside the local snippet. A no-bug judgment
that ignores those facts can miss executions that reach the
reported error state.

\subsection{A Harness Supplies the Analysis Context}
\label{sec:motivation-harness}

The bug above is not beyond program analysis. Given the relevant
code and inputs, a sufficiently precise static analyzer such as
\framac/\eva~\cite{cuoq2012framac} can track the full range of
the input through the guard and narrowing conversion, and check
the resulting array bound. In this case, the guard fails to
exclude values that cause an out-of-bounds write.

The challenge is maintaining such precision when analyzing a
real Linux kernel driver. Starting from module initialization
may require analyzing substantial driver and framework state
and control flow, making precise analysis expensive or
intractable. Conversely, analyzing the reported operation
in isolation may lose the constraints established earlier
in the program.
The analyzer needs something in between: the context the
warning depends on, without the rest of the driver.

A warning-specific harness supplies the 
analysis context. It
initializes the driver and framework state needed to reach the
reported operation, invokes a selected entry point, and represents
the inputs relevant to the warning abstractly.
This context could be modeled manually or generated by
subsystem-specific rules, but doing so across kernel subsystems
is costly. \tool instead uses an LLM to construct the harness
from the surrounding driver and framework code; \eva then
determines whether the reported error can occur.

Whether LLM-generated harnesses can provide sufficient context
for \eva to filter out false alarms in real kernel drivers remains
an empirical question. In \S\ref{sec:eval-dismissals}, we examine
how often this approach succeeds across 200 warnings and why
it fails.

\subsection{The Need for Harness Validation}

Keeping the final decision with the analyzer avoids relying on the LLM's
judgment, but the analyzer checks the generated harness, not the original
program. If a bug is reachable in the original program, it must also be
reachable in the harness for a no-bug result to justify filtering out
the warning. However, the LLM may inadvertently restrict relevant inputs
or exclude execution paths that trigger the bug, causing the analyzer
to find no bug even when one exists.

In the motivating example, the LLM could infer that \texttt{ucontrol->value[0]}
normally falls between 0 and 16 and encode this plausible constraint directly:

\begin{lstlisting}[language=C]
ucontrol->value[0] = Frama_C_interval(0, 16);
\end{lstlisting}

The range is reasonable for normal use of the driver, but no such constraint
exists on the attacker-controlled input in the original program (our threat model).
 The harness
therefore excludes the negative values that trigger the reported error. \eva can
then correctly conclude that the error is unreachable in the generated harness,
even though that conclusion does not hold for the original program.

This example shows why externalizing the LLM's assumptions as code is useful
but not sufficient. The assumption is now explicit, so \tool can check
properties of the generated harness that warning-relevant behavior depends
on. Section~\ref{sec:design-validation} motivates each check with the
construction error it targets, starting from this one.

\section{System Overview}
\label{sec:overview}

Figure~\ref{fig:motivation-overview} also summarizes \tool. \tool starts from a
warning produced by a front-end detector. The warning identifies an untrusted
source, a reported error state, a value-flow witness, and a call witness. From
the warning and the surrounding program, the LLM constructs a smaller C program
that sets up the selected analysis context. We call this program a
\emph{warning-specific analysis harness} and denote it by \(H\). \tool checks
properties of \(H\) on which a no-bug result depends and, if the harness is
accepted, passes it to a
backend analyzer to determine whether the reported error is reachable. In our
implementation, the backend is \framac/\eva~\cite{blazy2017eva}. \framac is a
program-analysis platform for C; \eva statically over-approximates the possible
values and program states to check whether runtime errors may occur. The design
is not tied to a particular front-end detector, provided that its warning can be
converted into the representation below.

\textbf{Program and error state.}
Fix a build of the analyzed program \(P\). A \emph{location} \(\ell\) is a
statement in that build, and the \emph{operation} at \(\ell\) is the memory
access or computation that the statement performs. An \emph{error state} is a
pair \(e=(t,\varphi)\), where \(t\) is the reported sink and \(\varphi\) is a
predicate over the program state at \(t\). An execution \emph{reaches} \(e\) if
it visits \(t\) in a state satisfying \(\varphi\); \(e\) is \emph{reachable} in
a program if some execution of that program reaches it.

\textbf{Warning.}
\tool represents a warning as \(W=(s,\pi_v,\pi_f,e)\). The source \(s\) is the
value or object that the detector marks as untrusted. It need not be the scalar
that later appears in the sink expression; in the motivating example, \(s\) is
the control object \texttt{*ucontrol}. The two witnesses describe the report at
two levels.

The \emph{value-flow witness} \(\pi_v=\langle p_0,p_1,\ldots,p_k\rangle\) is a
sequence of \emph{propagation points}: locations at which, according to the
detector, a value derived from \(s\) is read, transformed, or stored on its way
to the sink. The witness starts at the source and ends at the sink. We identify
both with their locations and write \(p_0=s\) and \(p_k=t\); the points
\(p_1,\ldots,p_{k-1}\) are the \emph{intermediate} points of the witness. For
the warning in Figure~\ref{fig:motivation-overview}, the detector reports
\(\texttt{*ucontrol}\leadsto\texttt{n}\leadsto\texttt{i}\leadsto\texttt{items[i]}\),
a witness with two intermediate points. Propagation points may lie in several
functions.

The \emph{call witness} \(\pi_f=\langle f_0,\ldots,f_m\rangle\) is the sequence
of functions through which the detector reports that execution reaches the sink function \(f_m\), which contains \(t\). It records function-level
transitions, including indirect calls through function pointers, but not the
objects or callback bindings that make those transitions possible. An analysis
that enters \(P\) at \(f_i\) cannot visit the propagation points reported in
the skipped prefix \(f_0,\ldots,f_{i-1}\); the remaining points of \(\pi_v\)
are the points \emph{relevant} to that entry.

Different detectors report this information in different forms; it is converted
into this representation before the LLM is invoked. Both witnesses describe the
detector's report, not established program behavior. They tell \tool which
computation and which call path the harness must exercise; they are not
treated as evidence that \(e\) is reachable.

The question is whether \(e\) is reachable in the analyzed program \(P\).
Filtering out the warning asserts that no execution of \(P\) reaches \(e\).

\textbf{Scope.}
\tool targets warnings in kernel drivers and similar code, where analysis
requires context from callers, framework state, or the surrounding environment.
A warning-specific harness reconstructs this context and provides an entry
point for the backend. We study kernel drivers because
this gap is especially large: relevant context often comes from build
configuration, framework state, and subsystem dispatch. The same problem also
arises in libraries, event handlers, and other mid-program entry points,
although we do not evaluate those settings here.

Our design assumes that the front-end report provides the untrusted input,
a value-flow witness, a call witness for identifying candidate entries,
and a checkable error state at the reported sink. The warnings studied in this
paper are taint-style memory- and integer-safety warnings. The same structure
can apply to other dataflow-style warnings that provide this information;
warnings with a different structure would require different information from
the detector and different harness checks. \tool treats values marked by the
warning or source specification as attacker-controlled, but does not determine
end-to-end exploitability.

\textbf{Outcomes.}
\tool produces three outcomes: \emph{not\_a\_bug}, \emph{still\_a\_bug}, and
\emph{unresolved}. It returns \emph{not\_a\_bug} only when the generated harness
passes all validation checks and the backend establishes that the reported error
is unreachable in \(H\). If the harness passes validation but the backend does
not establish unreachability, \tool returns \emph{still\_a\_bug}. This outcome
does not assert that the warning is a confirmed bug; it means that the completed
analysis could not rule out the reported error. If the pipeline cannot obtain an
accepted harness or a usable backend result, \tool returns
\emph{unresolved}.

Only \emph{not\_a\_bug} filters out the warning. Both \emph{still\_a\_bug} and
\emph{unresolved} keep it for inspection: the former means that backend
analysis completed but could not rule out the error, whereas the latter means
that the pipeline did not produce a usable analysis result.

\textbf{Checkable artifacts.}
A completed analysis records the generated harness \(H\), the result of each
validation check, and the backend analyzer's result for \(e\) on \(H\). These
artifacts can be inspected and rerun independently of the LLM's textual
reasoning. For a \emph{not\_a\_bug} outcome, they provide evidence for filtering out
the warning.

\textbf{Harness validation.}
The backend analyzes \(H\), not \(P\). For a no-bug result on \(H\) to carry
over to \(P\), the harness must not have made the reported error state
unreachable. The validation checks do not prove that the reported error state
remains reachable in \(H\) whenever it is reachable in \(P\). They target common
harness errors that can hide the reported error state, such as
constraining an input beyond what \(P\) allows or leaving part
of the reported witness unexercised. A failed check provides concrete feedback
for repair. In our evaluation across the tested LLMs, we did not observe
an accepted harness that exhibited the construction errors targeted by these
checks. Section~\ref{sec:design-validation} describes the checks and
Section~\ref{sec:discussion} discusses their limitations.

\section{Design}
\label{sec:design}

\begin{table}[t]
\centering
\begin{minipage}[t]{0.50\linewidth}
\vspace{0pt}
\setlength{\tabcolsep}{3pt}
\renewcommand{\arraystretch}{0.97}
\captionsetup[table]{font=footnotesize,labelfont=rm,textfont=rm,skip=3pt}
\captionof{table}{Warning information for the motivating example.}
\scalebox{0.80}{%
\begin{tabularx}{1.25\linewidth}{@{}l@{\hspace{5pt}}X@{}}
\toprule
\textbf{Item} & \textbf{Instance in Fig.~\ref{fig:motivation-code}} \\
\midrule

\rowcolor{black!5}[0pt][0pt]
\multicolumn{2}{@{}l@{}}{\textbf{\textit{Direct bug detector output}}} \\

Source \(s\) &
\texttt{*ucontrol} \\

Sink \(t\) &
write to \texttt{items[i]} \\


Error state \(e\) &
\((t,\texttt{i}\ge\texttt{ARRAY\_SIZE(items)})\) \\

Sink function \(f_m\) &
\texttt{msm\_routing\_put\_...} \\
\midrule
\rowcolor{black!5}[0pt][0pt]
\multicolumn{2}{@{}l@{}}{\textbf{\textit{Deterministic derived (no LLM)}}} \\

Value-flow witness \(\pi_v\) &
\(\pi_v=\langle p_0=s,\,p_1,\ldots,p_k=t\rangle\) \\

Source function \(f_0\) &
\texttt{snd\_ctl\_ioctl} \\

Call witness \(\pi_f\) &
\(\langle f_0,\ldots,f_m\rangle\) \\

Entry candidates \(E\) &
\(E=\{\pi_f[i..m] \mid 0 \le i \le m\}\) \\

\bottomrule
\end{tabularx}%
}
\label{tab:harness-spec-example}
\end{minipage}\hfill
\begin{minipage}[t]{0.47\linewidth}
\vspace{0pt}
\setlength{\intextsep}{0pt}
\captionsetup*{type=algorithm}
\captionsetup[algorithm]{font=footnotesize,labelfont={rm,bf},textfont=rm,labelsep=space,skip=2pt}
\algrenewcommand\algorithmicindent{0.8em}
\begin{algorithm}[H]
\fontsize{7.5}{9.2}\selectfont
\caption{\tool Bug Analysis Pipeline}
\label{alg:tool-refinement}
\begin{algorithmic}[1]
\Require Warning $W$
\Ensure \texttt{not\_a\_bug}, \texttt{still\_a\_bug}, or \texttt{unresolved}

\State $\mathit{diag} \gets \emptyset$
\For{$i \gets 1$ \textbf{to} $N_{\max}$}
    \State $H \gets \Call{LLMConstructHarness}{W,\, \mathit{diag}}$
    \State \textbf{if} $H = \bot$ \textbf{then} \Return \texttt{unresolved}
    \If{\textbf{not} \Call{ValidateHarness}{$H,\, W}$}
        \State $\mathit{diag} \gets \Call{ValidationDiagnostics}{H,\, W}$
        \State \textbf{continue}
    \EndIf
    \State \Return \Call{AnalyzeAcceptedHarness}{$H,\, W$}
\EndFor
\State \Return \texttt{unresolved}
\end{algorithmic}
\end{algorithm}
\end{minipage}
\end{table}

Algorithm~\ref{alg:tool-refinement} shows \tool's workflow at a high level.
Given the warning \(W\) defined in \S\ref{sec:overview}, \tool asks the LLM to
construct a candidate harness \(H\). The value-flow witness identifies the
reported computation that the harness should exercise, while the call witness
provides candidate entry points. Both witnesses come from the front-end report
and remain fixed throughout harness construction; the LLM neither produces nor
revises them. For the motivating warning, this means constructing a context in
which the reported out-of-bounds access at \texttt{items[i]} can be analyzed
while the value derived from \texttt{ucontrol} remains unconstrained within its
\texttt{long} type.

\tool compiles the generated harness and checks it before backend analysis. The
checks target recurring failure modes, such as narrowing inputs relevant to the
warning or failing to preserve warning-relevant locations reported by the
detector. If a check fails, its result is returned to the LLM
for a repair attempt. The LLM may also report that it finds no valid repair
(\(H=\bot\) in Algorithm~\ref{alg:tool-refinement}), in which case the warning
remains unresolved. A harness that passes validation proceeds to the backend,
which analyzes the reported error state and produces one of the outcomes defined
in \S\ref{sec:overview}.

\begin{figure}[t]
\centering
\begin{minipage}[t]{0.48\linewidth}
\vspace{0pt}
\input{arxiv-code/listing-01.tex}
\end{minipage}\hfill
\begin{minipage}[t]{0.50\linewidth}
\vspace{0pt}
\input{arxiv-code/listing-02.tex}
\end{minipage}
\caption{The last two functions of the call witness for the warning in
Figure~\ref{fig:motivation-overview}, simplified.}
\label{fig:motivation-code}
\end{figure}

\subsection{Warning Witnesses and Entry Candidates}
\label{sec:design-warning-input}

The two witnesses in \(W\) serve different purposes during harness construction.
The value-flow witness identifies the reported computation that the harness
should exercise, while the call witness identifies ways to enter that
computation. Both come from the warning information defined in
\S\ref{sec:overview}; \tool only reorganizes them for harness construction.

\textbf{Value-flow witness.}
The value-flow witness \(\pi_v\) records the program locations through which the
warning-dependent value is reported to propagate. These locations may span
multiple functions. For harness construction, \tool groups them by function and
uses \(\pi_v^{f_m}\) to denote the portion inside the sink function \(f_m\).

In the motivating example, \(f_m\) is
\texttt{msm\_routing...}. Within this function, the reported value moves
from \texttt{*ucontrol} to \texttt{n}, then to the loop index \texttt{i}, and
finally reaches the write to \texttt{items[i]}:
\[
  \texttt{*ucontrol}
  \leadsto \texttt{n}
  \leadsto \texttt{i}
  \leadsto \texttt{items[i]}.
\]
These reported locations tell the LLM which computation must remain present in
the generated harness. They also provide the propagation points whose reachability is checked by the
validation in \S\ref{sec:design-validation}.

\textbf{Call witness.}
The call witness
\(\pi_f=\langle f_0,\ldots,f_m\rangle\) records how execution reaches the sink function. In Figure~\ref{fig:motivation-code},
\(f_0\) is \texttt{snd\_ctl\_ioctl},
\(f_2\) is \texttt{snd\_ctl\_elem\_write}, and
\(f_3\) is \texttt{msm\_routing\_put\_...}; \(f_1\), omitted from the figure for
clarity, is another ALSA framework routine.

\tool deterministically derives candidate entries by taking suffixes of this
reported call witness:
\[
  E=\{\pi_f[i..m]\mid 0\leq i\leq m\}.
\]
Each suffix represents a different depth at which the harness can enter the
reported execution. Starting near \(f_0\) preserves more of the framework path,
but requires the harness to initialize more kernel state. Starting closer to
\(f_m\) avoids much of that framework code, but requires the harness to recreate
the state that the skipped calls would otherwise establish. Checks that
the skipped calls perform on the input are skipped with them. The analysis
then admits inputs that the program would reject, which can keep a false
alarm but cannot hide a bug.

The call witness can also contain indirect framework calls. In the motivating
example, \texttt{snd\_ctl\_elem\_write} calls the driver callback \texttt{msm\_routing\_put\_...} 
through
\texttt{kctl->put(...)}. The witness reports this transition, but does
not provide the concrete object wiring needed to execute it. When constructing
the harness, the LLM inspects the source to recreate the relevant control object
and bind its \texttt{put} operation to the reported callback.

\begin{figure}[t]
\centering
\includegraphics[width=\linewidth]{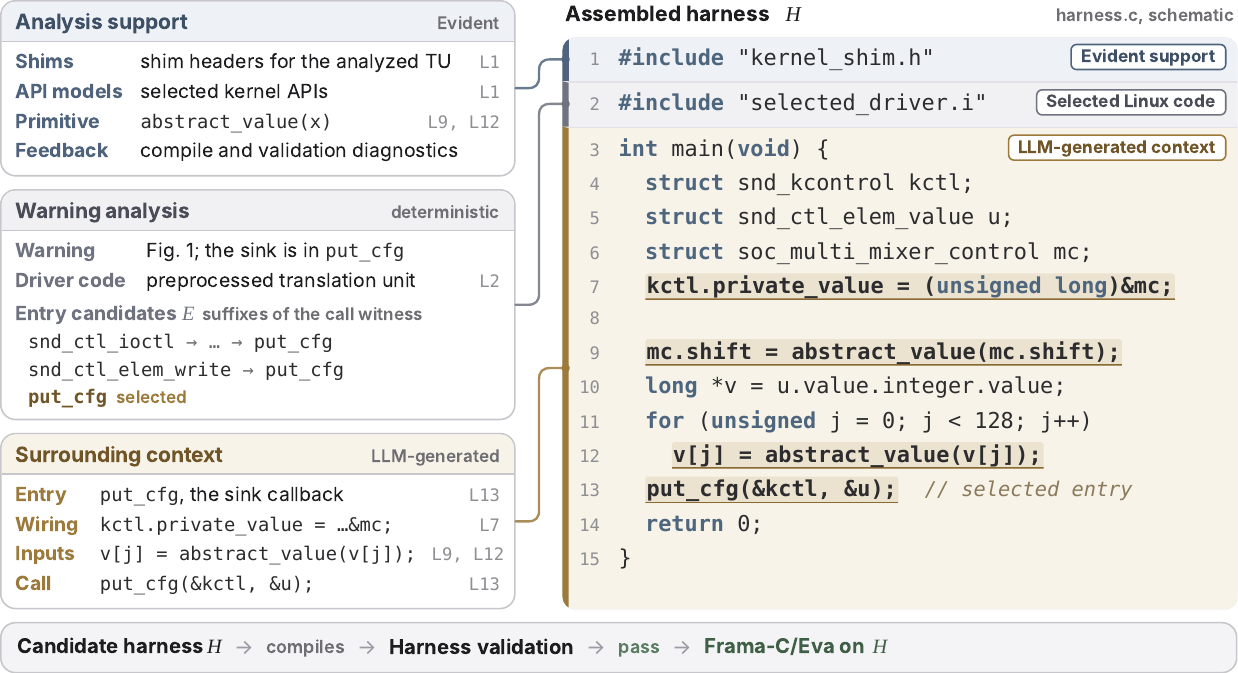}
\caption{Harness construction for the motivating warning: reused Linux code,
\tool-provided analysis support, and LLM-generated surrounding context.
The entry is the sink callback, abbreviated as \texttt{put\_cfg}; file names
are schematic.}
\label{fig:harness-sketch}
\Description{The harness combines a preprocessed Linux translation unit,
Evident's shim and abstract-value API, and LLM-generated object wiring,
abstract-input initialization, and an entry call. Compilation or validation
failures return diagnostics to the LLM for repair. Accepted harnesses proceed
to Eva.}
\end{figure}

\subsection{LLM-Guided Harness Construction}
\label{sec:design-harness-synthesis}

\textbf{Task.}
The harness constructor is the only component in \tool that uses an LLM. Given
the warning information and entry candidates from
\S\ref{sec:design-warning-input}, the LLM constructs a candidate C harness \(H\)
for one suffix \(\pi_f[i..m]\). It must construct the arguments and state needed
to invoke \(f_i\), initialize the numeric inputs relevant to the warning according to the
threat model below, and exercise the reported computation and call path to the sink function \(f_m\).

We assume adversarial user inputs and an uncompromised kernel. The harness
must leave attacker-controlled numeric inputs unconstrained within their C
types while constructing a valid kernel execution context. The LLM decides the structure of the context: which entry to call, which
objects to build, and how to connect them. Object identities, pointer
connections, callback bindings, and string literals are therefore concrete.
Any numeric input written directly by generated harness code should range over
its full C type; kernel execution may subsequently establish or constrain those
values through assignments, guards, and control flow.

For the motivating warning, this task is not just to call the driver callback.
If the harness starts from the ALSA framework routine
\texttt{snd\_ctl\_elem\_write}, it must construct the control objects needed for
the framework lookup and bind the driver's \texttt{put} operation so that the
indirect call reaches \texttt{msm\_routing\_put\_...}. If the harness starts
closer to the driver callback, it must instead reconstruct the objects that the
framework would have supplied, including the \texttt{kcontrol} and
\texttt{ucontrol} arguments. The entry candidates leave this choice to the LLM.

\textbf{Harness composition and the LLM's additions.}
Figure~\ref{fig:harness-sketch} separates the assembled harness into three
parts. First, \tool reuses the selected Linux code through a preprocessed
translation unit (TU), which contains the driver code and declarations under
the kernel's build configuration. Second, \tool supplies the analysis support:
shim headers, selected kernel API models, and the
\texttt{abstract\_value} primitive. Third, the LLM generates the surrounding
execution context: it selects an entry from the call-witness suffixes,
declares and wires the required objects, places abstract-value initializations,
and invokes the selected entry. The TU preparation and support layer are
described in \S\ref{sec:implementation}.

In the figure, the selected entry is the sink callback, abbreviated as
\texttt{put\_cfg}. The LLM constructs the control objects, connects
\texttt{kctl.private\_value} to \texttt{mc}, and initializes
\texttt{mc.shift} and the elements of \texttt{u.value.integer.value} using
\texttt{abstract\_value}. The local pointer \texttt{v} retains the array
elements' \texttt{long} type; the loop bound 128 denotes the array capacity,
not a restriction on input values. The include directives are schematic: the
implementation combines the preprocessed translation unit with the generated
harness code. Validation checks this program through the analyzer without
modifying it. Compilation or validation failures provide diagnostics for the
repair loop described below.


\textbf{Type-preserving abstract-value initialization.}
For warning-relevant numeric inputs, \tool uses \emph{type-preserving
abstract-value initialization}. Instead of asking the LLM to choose a concrete
range, the harness initializes each input with an abstract value whose domain is
determined by the C type of the lvalue:
\begin{lstlisting}[language=C]
x = abstract_value(x);
\end{lstlisting}
Here, \(x\) is the numeric lvalue being initialized, and
\texttt{abstract\_value} is a harness primitive that introduces an Eva abstract
value over the full domain of \(x\)'s C type. The LLM chooses where the input is
introduced; the program type determines its value range. The primitive applies
only to numeric values. Pointers and object identities are instead established
through valid storage and object wiring, as illustrated by
\texttt{kctl.private\_value} and \texttt{v} in
Figure~\ref{fig:harness-sketch}.

This distinction matters in kernel code, where warning-relevant values often
appear through typedefs, nested structures, and unions. In the motivating
warning, the relevant union field has type \texttt{long}, although its value is
later stored in an \texttt{int}. Asking the LLM to construct a backend-specific
interval would require it to choose both the type and the range, creating an
opportunity to exclude attacker-controlled values. With
\texttt{abstract\_value}, the LLM only identifies the lvalue to initialize;
the value range follows from its program type.

\textbf{Code navigation.}
Harness construction follows the pattern of tool-augmented coding agents: the
LLM is not expected to infer the relevant driver and framework structure from a
single prompt. \tool provides the warning input, entry candidates, relevant
snippets, and abstract-input requirements, and lets the LLM look up code
definitions in the kernel source. In our implementation, the LLM requests
definitions for a symbol or use site through \texttt{get\_definition}, backed
by a lightweight tool layer using clangd~\cite{clangd,lsp}. The returned
definitions provide the source declarations and code needed to construct the
context.

\textbf{Repair loop.}
Harness construction is iterative. After each proposal, \tool builds the
harness and runs the available validation checks. Any failure is returned to
the LLM as a diagnostic, and the LLM is asked to revise the harness for the
same warning. Failures may be ordinary compile errors, such as missing
declarations, incompatible pointer types, or invalid field accesses, or
validation failures, such as narrowing an abstract numeric input before the
entry call, failing to reach the reported sink, or leaving a reported
intermediate propagation point unreachable. We describe these checks in more detail
in \S\ref{sec:design-validation}.

A validation failure does not necessarily mean that a valid repair exists.
The LLM is therefore asked to preserve the intended harness and its existing
constraints rather than bypass a failed check. Repair is bounded, and the LLM
may also report that it cannot find a valid repair under these constraints.
If no candidate passes validation within the repair budget, or the LLM reports
that no valid repair is available, \tool marks the case unresolved.

\subsection{Harness Validation}
\label{sec:design-validation}

The LLM is asked to construct the context in which the reported
computation can be analyzed, leaving the bug decision to \eva.
We assume that the LLM attempts to follow this task; deliberate attempts
to evade validation are outside our scope. Construction errors,
however, may restrict inputs that trigger a real bug or prevent \eva
from analyzing the reported code, causing the bug to be filtered out (FN).

Validation does not attempt to establish that the harness preserves
every behavior relevant to the warning.
We derive the requirement behind each check from the original program
and the detector report, and use \eva's analysis of the generated
harness to check it. These checks do not rely on assumptions introduced by the
LLM or require \eva to produce a particular bug result.

Because the warnings we consider come from taint-style analyses, we
focus validation on the reported value flow. The detector report
identifies the tainted source and the program locations along the
reported flow from the source to the sink. We therefore check whether
the harness restricts the tainted inputs, which we conservatively take
to be the numeric values it supplies to the entry, and whether \eva
analyzes the reported flow. For the latter, the report provides program
locations but not a representation of value dependencies that \eva can
check directly. We thus approximate validation of the reported value
flow through program-point reachability.

We implement three validation checks.
\textbf{Input range} checks whether tainted numeric inputs retain their
full type ranges, rather than being restricted to values the LLM
considers plausible.
\textbf{Sink reachability} checks whether \eva reaches and analyzes the
reported sink operation.
\textbf{Path reachability} checks for gaps along the reported flow,
where an intermediate location is unreachable even though a later
reported location is reached.
Table~\ref{tab:harness-validation} summarizes the three checks.

\begin{table}[t]
\centering
\caption{Harness validation checks
(\S\ref{sec:check-input}--\S\ref{sec:check-path}). Instances are from our
runs.}
\label{tab:harness-validation}
\footnotesize
\renewcommand{\arraystretch}{1.08}
\begin{tabularx}{\linewidth}{@{}l>{\raggedright\arraybackslash\hsize=.88\hsize}X>{\raggedright\arraybackslash\hsize=1.14\hsize}X>{\raggedright\arraybackslash\hsize=.98\hsize}X@{}}
\toprule
\textbf{Check} & \textbf{Construction error} & \textbf{Instance} & \textbf{Observed behavior} \\
\midrule
Input range &
Narrows a tainted input before the entry call &
\mbox{\texttt{if (idx >= MAX) return 0;}} before the driver call &
Value loses range before the entry call \\
\addlinespace[2pt]
Sink reachability &
Stops the analysis before the sink &
\mbox{\texttt{struct \{ int shift; \} model;}} with the wrong layout &
No abstract state reaches the sink operation \\
\addlinespace[2pt]
Path reachability &
Skips an intermediate point, still reaches the sink &
\mbox{\texttt{algo.smbus\_xfer = NULL;}} with a fallback path &
Intermediate point unreachable, later point reachable \\
\bottomrule
\end{tabularx}
\end{table}

Failed checks are returned to the LLM as repair feedback.
If validation still fails after repair, the warning is left unresolved.

\subsubsection{Input Range}
\label{sec:check-input}

Type-preserving initialization prevents the LLM from choosing an overly narrow
range when a numeric input is introduced. It does not, however, prevent
generated setup code from narrowing that value afterwards. In one generated
harness, the LLM correctly initialized an index with
\texttt{abstract\_value}, but then inserted a bounds check before calling the
driver entry:
\begin{lstlisting}[language=C]
ucontrol.id.index = abstract_value(ucontrol.id.index);
if (ucontrol.id.index >= MAX_CHANNELS)
    return 0;
snd_us16x08_route_put(&kcontrol, &ucontrol);
\end{lstlisting}
The initialization is correct, but the added check removes the out-of-range
values relevant to the warning before the original program is entered.

\textbf{Assignment check.} 
The first check operates on the typed AST of the generated code. Assignments to
numeric inputs must introduce an abstract value over the full C type, rather
than a constant or a range chosen by the LLM. Structural operations such as
pointer and callback wiring are not subject to this requirement.

\textbf{Entry check.}
The second check uses the abstract values computed by \eva to determine whether
generated setup code subsequently narrows an input. For each numeric input
relevant to the warning, we compare the value of the same storage immediately after its
initialization (\(A\)) and immediately before execution first enters the
selected entry of the original program (\(B\)). \(A\) must cover the full
domain of the storage type, and \(B\) must contain every value represented at
\(A\). Any loss between the two points indicates that generated setup code has
restricted the input. Constraints introduced inside the original program occur
after \(B\) and are therefore not treated as harness restrictions.

\textbf{Scope.}
Both checks target inadvertent narrowing of the numeric inputs
relevant to the warning during harness construction. It does not cover more complex
data-dependent or relational constraints, nor deliberate attempts to circumvent
the check.

\subsubsection{Sink Reachability}
\label{sec:check-sink}

A harness may preserve the numeric inputs correctly but still fail to provide
the context needed to analyze the reported operation. In one generated
harness, the LLM constructed the object carrying the input with the wrong
layout:
\begin{lstlisting}[language=C]
/* generated harness */
struct { int shift; } model;
model.shift = abstract_value(model.shift);
kcontrol.private_value = (unsigned long)&model;

/* original driver */
int shift = ((struct soc_multi_mixer_control *)
             kcontrol->private_value)->shift;
\end{lstlisting}
In the real structure, \texttt{shift} is preceded by other fields. The driver's
cast therefore reads beyond the bounds of the smaller generated object. \eva
reports a definite invalid access and loses all states before reaching the
reported operation. The analysis itself terminates normally, but the absence
of an alarm at the sink says nothing about the warning because the sink
was never analyzed.

\textbf{Sink reachability.}
\tool therefore checks whether \eva reaches and evaluates the reported
sink operation, not whether the analysis can proceed beyond it.
If all states are lost before the sink, the check fails.
If \eva reports the sink operation itself as invalid, it still
counts as reached, even when no valid successor states remain.

An unreachable sink can be benign if it is unreachable
in the original program regardless of the inputs chosen by the
harness (\eg dead code).
In contrast, a sink made unreachable by the generated harness
is not benign. Neither is an analysis that stops because of an
error before the sink or an analysis failure.

\textbf{Scope.}
Sink reachability does not require every execution to reach the reported
operation. The original program may legitimately reject some inputs or leave
unused object fields uninitialized. The check only asks whether the harness
permits the reported operation itself to be analyzed; failures on
unrelated branches are not validation failures for this warning.

\subsubsection{Path Reachability}
\label{sec:check-path}

Sink reachability alone does not show that the harness preserves the value-flow
witness reported by the detector. In one generated harness, the LLM left a
transfer callback unconnected:
\begin{lstlisting}[language=C]
/* generated harness */
algo.smbus_xfer = NULL;
adapter.algo = &algo;

/* original driver, simplified */
if (adapter->algo->smbus_xfer)
    adapter->algo->smbus_xfer(...);  /* reported propagation point */

i2c_smbus_xfer_emulated(...);        /* fallback */
\end{lstlisting}
The null callback skips the reported transfer path, but execution continues
through the fallback and later reaches the reported sink. The harness
therefore passes both input range and sink reachability while bypassing
part of the reported value flow.

\textbf{Propagation gaps.}
Not every unreachable location in the detector-reported value-flow
witness indicates a problem with the harness. For example, the
original program may return normally, leaving the remaining
locations unreachable. We instead look for a more specific pattern:
an intermediate location is unreachable, but a later reported
location remains reachable. Such a gap suggests that the harness
may fail to exercise part of the reported computation, even though
the analysis can still reach downstream code. The null callback
above produces exactly this pattern.

\textbf{Reachability check.}
For the selected entry, \tool queries the reachability of the relevant
propagation points in \(\pi_v\) using \eva. A propagation point is reachable
when an abstract state reaches the location immediately before its operation.
The check considers these results in witness order together with the
termination information from sink reachability. A missing suffix explained by
a normal return from the original program is not by itself a validation
failure. If an intermediate point is unreachable while a later point in
\(\pi_v\) remains reachable, \tool reports a path-reachability failure and
returns it for repair.

\textbf{Scope.}
The check targets a harness that makes an intermediate reported
location unreachable while later locations remain reachable. It checks
reachability of the reported locations, but does not establish that one
execution visits them in order or that the reported value propagates through
them.

\subsection{Analyzing the Harness}
\label{sec:design-decisions}

To filter out a warning, the analyzer must account for all possible
executions of the harness while being precise enough to rule out
the reported error.
We use \framac/\eva, a sound static analyzer based on abstract
interpretation.
\tool returns \emph{not\_a\_bug} only when \eva establishes
that the reported error cannot occur in the harness.

\textbf{Analysis outcomes.}
If \eva rules out the reported error, \tool returns
\emph{not\_a\_bug}. If the analysis completes but the error remains possible,
\tool returns \emph{still\_a\_bug}; this means that the warning could not be
filtered out, not that the bug has been proved feasible. If no usable analysis
result is obtained (\eg due to a timeout), \tool returns \emph{unresolved}. 
Thus, loss of precision
keeps the warning as \emph{still\_a\_bug}, whereas failure to obtain an analysis
result makes it \emph{unresolved}.


\textbf{Analysis cascade.}
More precise Eva configurations are more likely to rule out an infeasible
error, but they are also more expensive and may time out. \tool therefore
first seeks a completed analysis under its default quick configuration and
then increases precision when possible.

The quick analysis uses precision level 1 and Eva's default
\texttt{-eva-context-depth} of 2. If it times out, \tool retries the same
harness with \texttt{-eva-context-depth} reduced first to 1 and then to 0,
while keeping the remaining quick-analysis configuration unchanged. If none
of these runs completes, the case is \emph{unresolved}.

Once a quick analysis completes, \tool runs the full analysis at precision
level 11. If the full analysis completes, its result is used; if it times out,
the completed quick result is retained.

\section{Implementation}
\label{sec:implementation}

This section describes how \tool instantiates the design on Android kernel
driver code and \framac/\eva. We focus on the implementation details needed to
construct and analyze harnesses: warning import and kernel preparation, kernel
shims, harness construction, harness validation, and backend result
extraction.

\textbf{Warning Inputs and Kernel Build Preparation.}
\tool supports two forms of front-end warning input. The first is an
LLVM-based analysis, represented by SUTURE in our evaluation
\cite{DBLP:conf/ccs/ZhangCHLZZQ21}; its adapter imports the reported instruction
and call traces together with debug locations. The second is a static-analysis
report in SARIF format, represented by CodeQL-OOB in our evaluation, an existing
out-of-bounds detector built on CodeQL's taint-tracking library
\cite{backhouse_stack_2018}; its adapter imports the
reported source, sink, and value-flow locations. Both adapters
deterministically translate the front-end report into the warning
representation of \S\ref{sec:overview}.

For each analyzed driver file, we first build the kernel to recover its include
paths, configuration-dependent macros, and compiler options. The file is then
preprocessed into a macro-expanded \texttt{.i} translation unit. The generated
harness is combined with this
translation unit; validation and backend analysis operate on the resulting
program without inserting additional instrumentation.

\textbf{Harness Support and Kernel Shims.}
Generated harnesses are analyzed outside the full kernel environment. \tool
therefore provides shim headers, wrapper include paths, the
\texttt{abstract\_value} primitive, and models for selected kernel APIs. The
wrapper layer simplifies mechanisms that are either irrelevant to the warning
or difficult for C-level analysis, including compiler attributes, build-time
assertions, dynamic-debug macros, selected atomic primitives, and
configuration-specific helpers.

When a shim removes an operation from the analyzed program, it explicitly
marks that removal with a pragma that survives preprocessing. This lets later
analysis distinguish an operation removed by the analysis model from one that
is merely unreachable. Shims that return modeled values without removing an
operation do not use this marker.

The shim layer also implements type-preserving abstract-value initialization.
For a numeric lvalue \(x\), \texttt{abstract\_value(x)} uses the compile-time
type of \(x\) to select the corresponding \framac/\eva abstract expression and
casts the result back to \texttt{\_\_typeof\_\_(x)}. The generated harness
therefore chooses where an abstract input is introduced, while the C type
determines its value range.

Some warning classes require small contracts to expose the corresponding
property to \eva. For warnings on \texttt{copy\_from\_user}-style APIs, for
example, \tool models the destination-buffer validity obligation checked by
the warning:
\begin{Verbatim}[commandchars=\\\{\},numbers=left,xleftmargin=5pt,fontsize=\footnotesize,breaklines]
\PYG{c+cm}{/*@}
\PYG{c+cm}{requires n == 0 || \PYGZbs{}valid(((unsigned char *)to) + (0 .. n\PYGZhy{}1));}
\PYG{c+cm}{assigns ((unsigned char *)to)[0 .. n\PYGZhy{}1]}
\PYG{c+cm}{  \PYGZbs{}from ((const unsigned char *)from)[0 .. n\PYGZhy{}1], n;}
\PYG{c+cm}{assigns \PYGZbs{}result \PYGZbs{}from from, n;}
\PYG{c+cm}{ensures \PYGZbs{}result \PYGZlt{}= n;}
\PYG{c+cm}{*/}
\PYG{k+kt}{unsigned}\PYG{+w}{ }\PYG{k+kt}{long}\PYG{+w}{ }\PYG{n+nf}{copy\PYGZus{}from\PYGZus{}user}\PYG{p}{(}\PYG{k+kt}{void}\PYG{+w}{ }\PYG{o}{*}\PYG{n}{to}\PYG{p}{,}\PYG{+w}{ }\PYG{k}{const}\PYG{+w}{ }\PYG{k+kt}{void}\PYG{+w}{ }\PYG{o}{*}\PYG{n}{from}\PYG{p}{,}\PYG{+w}{ }\PYG{k+kt}{unsigned}\PYG{+w}{ }\PYG{k+kt}{long}\PYG{+w}{ }\PYG{n}{n}\PYG{p}{);}
\end{Verbatim}

\eva interprets the \texttt{requires} clause to check the precondition.
\tool filters out the warning only if \eva establishes that this precondition
holds.

Allocation routines such as \texttt{kmalloc} are mapped to \eva's allocation
builtins, which track the requested size, initialization, and lifetime of each
allocated object while allowing allocation to fail. Additional wrappers cover
kernel mechanisms such as \texttt{current},
\texttt{READ\_ONCE}/\texttt{WRITE\_ONCE}, common data-structure helpers, and
selected architecture- and configuration-specific idioms.

\textbf{Harnessing Harness Construction.}
Harness generation is implemented as a scaffolded, tool-augmented coding task.
The model receives the warning representation and the construction constraints
from \S\ref{sec:design-harness-synthesis}. \tool instructs the LLM not to redefine
existing kernel functions or tool-provided helpers. During planning, it can look
up code definitions through a code-navigation interface over the kernel source tree,
and then produces a plan identifying the entry, required object wiring, and
numeric inputs that must remain abstract. It subsequently emits the C harness
under that plan. Repair turns keep the warning fixed and add compiler or
validation diagnostics.

The scaffold enforces the separation between structural setup and numeric
inputs. Concrete pointer connections, callback bindings, object identities,
and string literals are permitted. Numeric values assigned by generated code,
including indices, lengths, flags, selectors, and counts, must instead use
\texttt{abstract\_value} over their full C type.

Whole-object initialization is rejected. In particular, generated code may
not zero an aggregate with \texttt{memset} or copy an aggregate wholesale with
\texttt{memcpy}. Such operations can silently fix fields that the model has
not inspected and thereby affect reachability or pointer validity. Harnesses
must initialize fields individually, using \texttt{abstract\_value} for numeric
inputs to avoid implicitly fixing them to zero. Missing fields can then surface
as concrete \eva diagnostics during repair.

\textbf{Input Range.}
The check is implemented in two steps. The first, the assignment check,
parses the harnessed program with \framac and inspects its typed AST. Numeric assignments introduced
by generated code must use \texttt{abstract\_value}; structural assignments are
distinguished using the typed AST rather than field names or model
explanations. Aggregate initialization is rejected at this stage as well.
Some concrete values are exempt, such as string literals, local loop
counters, and \texttt{sizeof} expressions over kernel objects. They serve the
wiring and orchestration that the harness is required to perform, rather than
acting as numeric inputs relevant to the warning.

The second, the entry check, checks the effect of the generated setup with a
separate, low-cost \eva run. For each abstract numeric input, \tool reads its abstract
value immediately after initialization (\(A\)) and again immediately before
the first call to the selected entry (\(B\)). \(A\) must cover the complete
domain of the input's C type, including the actual width of a bit-field, and
\(B\) must contain every value represented at \(A\). The comparison follows
the input itself rather than the pointer expression originally used to
initialize it.

\textbf{Sink Reachability and Path Reachability.}
Both are queried directly on the typed AST rather
than by inserting probes into the program. Preprocessing preserves
\texttt{\#line} information, which \tool uses together with the reported file,
line, and function to map the sink and the points in the value-flow witness
\(\pi_v\) to operations in the original program.

For a propagation point, \tool asks whether an abstract state reaches the
corresponding operation. Sink reachability additionally distinguishes what
happens at the reported operation. A sink passes if an abstract state
reaches the operation, whether or not evaluating it raises an alarm; an
alarm at the sink that leaves no valid successor state still counts as
reached. If every state is lost before the sink, for example by a definite
invalid access, validation fails and the blocking location is returned for
repair.

When the sink is not reached, \tool distinguishes a normal return through the
original program from an interruption caused by the harness. The
same information is used when interpreting path reachability. A normal
return can explain an unreachable suffix of \(\pi_v\); in contrast, an
unreachable intermediate propagation point followed by reachable downstream
points indicates that the harness has bypassed part of the reported
propagation. Validation queries use a fixed low-cost \eva configuration rather
than the backend cascade described below.

\textbf{Development of the Checks.}
We implemented and debugged the three validation checks on harnesses
generated by a single LLM (gpt-5.6-luna) for the same 200 warnings.
The checks were fixed before the evaluation in
\S\ref{sec:evaluation} and were not tuned to the harnesses of the
LLMs evaluated there.

\textbf{Frama-C/Eva Analysis and Result Extraction.}
Backend analysis begins with a quick \eva run at precision level 1 and the
default context depth of 2. If this run times out, \tool retries the same
harness with \texttt{-eva-context-depth 1}, and then with
\texttt{-eva-context-depth 0}. These quick runs use timeout budgets of 180,
300, and 180 seconds, respectively. If none completes, the case is
\emph{unresolved}.

Once a quick analysis completes, \tool runs the full analysis at precision
level 11 with a 900-second timeout. A completed full result takes precedence
over the quick result. If the full analysis times out, \tool retains the
completed quick result. Errors other than timeouts are handled as analysis
failures and do not trigger another stage of the cascade.

\tool extracts the backend result relative to the reported warning rather than
by counting all alarms emitted by \eva. The primary match is an alarm at the
detector-reported sink. To account for source-location drift introduced by
preprocessing and macro expansion, \tool also considers nearby reports in the
same enclosing function when they involve the same warning-dependent value and
the same class of error. For integer-overflow warnings, the effect may become
visible only in a later operation that depends on the value produced at the
reported sink; such dependent operations are also considered during result
matching. Operations explicitly removed by the shim model are excluded from
this matching.

A definite error before the reported sink is not treated as evidence for the
warning. It means that the analysis was interrupted before the reported
operation, and the case goes back to the LLM as a sink-reachability failure.

If \eva rules out the reported error on the accepted harness, \tool returns
\emph{not\_a\_bug}. If the analysis completes but the error remains possible,
the warning remains \emph{still\_a\_bug}. If no usable backend result is
available, \tool returns \emph{unresolved}.

\textbf{Implementation Size.}
The core implementation comprises approximately 18.8 KLOC of code.
It includes four \framac plug-ins in
OCaml (1 KLOC), the kernel environment models in C (0.6 KLOC), and the
prompt specifications (0.3 KLOC); the rest is Python. The validation
stages account for about 4.0 KLOC, 21\% of the total.
The deterministic
binding of indirect calls and selection of candidate entries takes 3.7 KLOC
(20\%).

\section{Evaluation}
\label{sec:evaluation}

We evaluate \tool on real warnings from Android Linux kernel drivers. We ask
three research questions:

\squishlist
\item \textbf{RQ1 (Practical Feasibility):}
To what extent can LLM-generated warning-specific harnesses
provide sufficient context for static analysis to filter out
false alarms in complex kernel drivers?

\item \textbf{RQ2 (Comparison):}
Compared with LLM-decided approaches, how many false alarms
and confirmed bugs does \tool filter out?

\item \textbf{RQ3 (Design Choices):}
How do removing harness validation and giving the LLM
more control affect how many false alarms and confirmed
bugs are filtered out?
\squishend

\subsection{Experimental Setup}

\textbf{Dataset.}
We evaluate on 200 warnings drawn from the BugLens dataset~\cite{buglens},
which includes Android Linux kernel driver warnings reported by two existing
static detectors: SUTURE~\cite{DBLP:conf/ccs/ZhangCHLZZQ21} and CodeQL-OOB,
a CodeQL-based out-of-bounds detector~\cite{backhouse_stack_2018}. The dataset
contains 126 SUTURE warnings and 74 CodeQL-OOB warnings. All three RQs use the
full set of 200 warnings.

For the SUTURE subset, we use the two main driver modules in the dataset,
\texttt{sound} and \texttt{i2c}, because prior BugLens results show that these
two modules cover all confirmed true bugs in that subset. The SUTURE benchmark
cases, \texttt{msm-sound} and \texttt{msm-i2c}, are evaluated on the Android 10
Qualcomm Pixel kernel \texttt{android-msm-coral-4.14-android10}
(\texttt{android-10.0.0\_r0.56}), Linux 4.14.150. The CodeQL-OOB cases use a
separate Qualcomm MSM kernel tree with Linux 4.4.21-perf+.

\textbf{Ground truth.}
Ground-truth labels were established from the labels and case analyses reported
in the SUTURE and BugLens papers, followed by additional manual validation by
the authors. In total, the ground-truth set contains 40 real bugs and 160
non-bugs. Of the 40 real bugs, 39 come from SUTURE and one from CodeQL-OOB.
We treat these labels as the best available reference for this dataset.

\textbf{Metrics.}
We treat a confirmed bug as the positive class and a false alarm as the
negative class. A false negative (FN) is a confirmed bug that a method
filters out; a false positive (FP) is a false alarm that is kept for inspection.
Unresolved warnings are kept and counted as TP or FP according to ground
truth, for both \tool and the baselines (Table~\ref{tab:eval-labels}). We
report the underlying counts, the false-negative and false-positive rates, and
accuracy:
\[
\mathrm{FNR} = \frac{FN}{TP+FN}, \qquad
\mathrm{FPR} = \frac{FP}{FP+TN}, \qquad
\mathrm{Acc} = \frac{TP+TN}{ALL}.
\]

\begin{table}[t]
\centering
\caption{Mapping \tool's outcomes to evaluation labels.}
\begin{tabular}{lll}
\toprule
\textbf{System result} & \textbf{Ground truth} &
\textbf{Evaluation label} \\
\midrule
\multicolumn{3}{>{\columncolor{gray!12}}l}{
  \textit{Explicit decision}} \\
\texttt{not\_a\_bug} & Non-bug & \textbf{TN} --- false alarm filtered out \\
\texttt{not\_a\_bug} & Real bug & \textbf{FN} --- real bug filtered out \\
\texttt{still\_a\_bug} & Real bug & \textbf{TP} --- real bug kept \\
\texttt{still\_a\_bug} & Non-bug & \textbf{FP} --- false alarm kept \\
\midrule
\multicolumn{3}{>{\columncolor{gray!12}}l}{
  \textit{No decision}} \\
\texttt{unresolved} & Any & \textbf{U} --- kept \\
\bottomrule
\end{tabular}
\label{tab:eval-labels}
\end{table}

\textbf{Implementation.}
In the primary configuration, \tool uses GPT-5.5
(\texttt{gpt-5\_5-2026-04-23}, reasoning effort \texttt{high}) for harness
construction and Frama-C 32.0 (Germanium), through the Eva plug-in, for static
analysis. In the full configuration, all three checks of
\S\ref{sec:design-validation} are enforced: a harness is accepted only if it
passes the input range, sink reachability, and path reachability checks.
The design variants in RQ3 selectively remove these constraints as described
there. In the main experiments, we run each method three times per warning
and report the majority outcome.

\textbf{Models.}
We evaluate the full \tool configuration with four LLMs: GPT-5.5,
GPT-5.4 mini, Claude Sonnet 4.6, and GLM-5.1~\cite{zai-orgglm-5_2026}.
The set includes two OpenAI models at different capability tiers and two
models from other model families. We use GPT-5.5 as the primary model:
unless stated otherwise, detailed single-model analyses use GPT-5.5, while
cross-model experiments report results for all four models.

\textbf{Cost and runtime.}
Cost is not a primary evaluation target. With
GPT-5.5, harness generation consumes about 27K uncached input tokens, 42K cached
input tokens, and 11K output tokens per case attempt, costing about \$0.50. This
is below the agentic Codex baseline in our runs (\$1.20/case). BugLens and
\tool use similar numbers of tokens, so their GPT-5.5 costs are similar.
Runtime is dominated by \framac/\eva.
Serial execution averages
about 10 minutes per case, but cases are independent and parallelize naturally
across workers.

\subsection{RQ1: Practical Feasibility}
\label{sec:eval-dismissals}

\begin{table}[t]
\centering
\caption{\tool with four LLMs on all 200 warnings. Kept counts completed
analyses; unresolved warnings are also kept.}
\label{tab:main}
\begin{tabular}{lrrrrrr}
\toprule
\multirow{2}{*}{\textbf{LLM}}
& \multicolumn{3}{c}{\textbf{False alarms (160)}}
& \multicolumn{3}{c}{\textbf{Confirmed bugs (40)}} \\
\cmidrule(lr){2-4}\cmidrule(l){5-7}
& \textbf{Filtered out} & \textbf{Kept} & \textbf{Unres.}
& \textbf{Filtered out} & \textbf{Kept} & \textbf{Unres.} \\
\midrule
GPT-5.5 (primary) & 102 & 57 & 1 & \textbf{0} & 40 & 0 \\
Claude Sonnet 4.6 & 98 & 49 & 13 & \textbf{0} & 39 & 1 \\
GLM-5.1           & 94 & 45 & 21 & \textbf{0} & 38 & 2 \\
GPT-5.4 mini      & 78 & 38 & 44 & \textbf{0} & 37 & 3 \\
\bottomrule
\end{tabular}
\end{table}

RQ1 evaluates whether \tool can filter out false alarms
without filtering out confirmed bugs.
We examine three questions:


\squishlist
\item \textbf{Does \tool filter out confirmed bugs?}
Errors in harness construction may cause \eva to miss a real bug.
We examine whether \tool incorrectly filters out any of the
40 confirmed bugs across the four LLMs.

\item \textbf{How many false alarms can \tool filter out?}
Keeping every warning trivially avoids false negatives,
but provides no filtering benefit.
We measure how many of the 160 false alarms \tool can filter out.

\item \textbf{Why do some false alarms remain?}
For false alarms that \tool cannot filter out, we investigate
whether the cause lies in harness construction, limitations
of \eva, or other factors.
\squishend


\subsubsection{Confirmed bugs}
Across all four LLMs, \tool does not filter out any of the 40 confirmed
bugs. With GPT-5.5, \tool keeps all 40 after completed analyses.
With Sonnet 4.6, GLM-5.1, and GPT-5.4 mini, it keeps 39, 38,
and 37, respectively, leaving six confirmed-bug outcomes unresolved
in total.

All six unresolved cases failed validation: 2 restricted
warning-relevant inputs, 3 failed sink reachability,
and 1 failed path reachability.
Despite receiving validation feedback, these weaker LLMs failed
to repair these harnesses.
\tool therefore left the warnings unresolved, keeping 
these cases for inspection.

\subsubsection{Filtering utility}
Across the four LLMs, \tool filters out 78--102 of the 160 false
alarms (48.8--63.8\%).
With GPT-5.5, our primary configuration, \tool filters out
102 false alarms (63.8\%) and keeps all 40 confirmed bugs.
Of the 200 warnings, 98 remain for inspection
(97 kept after analysis and one unresolved),
a reduction of 51.0\%.

The other LLMs filter out fewer false alarms and leave more
warnings unresolved: 14 with Sonnet 4.6, 23 with GLM-5.1,
and 47 with GPT-5.4 mini, compared with one for GPT-5.5.
Among false alarms with resolved outcomes, each model
filters out 64--68\%.
The models differ mainly in how many warnings they leave unresolved.

\subsubsection{Failure analysis}
With GPT-5.5, \tool kept 57 known false alarms and left one
unresolved.
We manually examined these 58 cases, comparing the original
code and its calling context with the generated harnesses,
validation feedback, and \eva's analysis results to identify
why the warnings were not filtered out.
We identified several recurring causes, which are not mutually
exclusive.

A common cause is \emph{incomplete program and environment
modeling}, which manifests in several ways.
First, \emph{inaccurate object construction} (4 FP) can
introduce errors absent from the original program.
For example, a table constructed in the harness may omit
its terminating entry, allowing traversal beyond its bounds.
Second, \emph{missing caller-side constraints} (9 FP) can
admit inputs that the original callers would reject.
For example, invoking a helper directly may bypass a
bounds check performed by its caller.
Third, \emph{incomplete modeling of called functions}
(21 FP) occurs when the analysis relies on function models
or stubs instead of the callees' implementations.
These models may omit constraints established by the actual
functions, leaving subsequent operations potentially unsafe.

Another recurring cause is \emph{limited static-analysis
precision} (22 FP).
In these cases, the relevant safety checks or relationships
are present in the analyzed code, but \eva cannot establish
that they rule out the reported error.

Finally, SUTURE and \eva use different internal representations
of the same program (LLVM IR and Frama-C's CIL-based AST, respectively).
Source locations do not always identify the same operation
across these representations, particularly after macro expansion.
This can cause incorrect alarm attribution or leave the reported
sink unidentified, as happened in GPT-5.5's single unresolved
false alarm.

\begin{takeaway}{RQ1}
An LLM-built, warning-specific harness is often enough to filter out a false
alarm: with GPT-5.5, \eva filters out 63.8\% of the 160 false alarms on such
harnesses and none of the 40 confirmed bugs.
\end{takeaway}

\subsection{RQ2: Safety--Utility Tradeoff}
\label{sec:eval-tradeoff}

RQ1 establishes \tool's effectiveness, but does not show
how its results compare with existing LLM-based approaches.
Allowing LLMs to make the final bug decision may filter out
more false alarms, but also risks filtering out confirmed bugs.
RQ2 examines this tradeoff by comparing how many false alarms
and confirmed bugs each method filters out.

We compare \tool with BugLens~\cite{buglens}, an existing
LLM-based system for filtering static analysis warnings,
across different LLMs, and Codex, a general-purpose coding agent.
Both BugLens and Codex leave the final bug decision to the LLM.
We also evaluate an FN-aware variant of Codex to examine
whether explicitly instructing the agent to avoid false negatives
can reduce this risk.

\begin{table}[t]
\centering
\caption{Safety and utility of \tool and LLM-decided baselines.
Unresolved or uncertain warnings are kept; parentheses give the confirmed
bugs among them.}
\label{tab:llm-filtering}
\small
\begin{tabular}{llrrrrrr}
\toprule
& & \multicolumn{2}{c}{\textbf{Filtered out}} & &
\multicolumn{3}{c}{\textbf{Rates}} \\
\cmidrule(lr){3-4} \cmidrule(lr){6-8}
\textbf{LLM} & \textbf{Method}
& \textbf{Non-bugs} & \textbf{Bugs}
& \textbf{Unres.} & \textbf{FPR} & \textbf{FNR} & \textbf{Acc (\%)} \\
\midrule

 & Codex            & 124 & 18 & 1 (1) & 0.23 & 0.45 & 73.0 \\
 & Codex (FN-aware) & 123 & 11 & 4 (4) & 0.23 & 0.28 & 76.0 \\
 & BugLens          & 127 & 14 & 0     & 0.21 & 0.35 & 76.5 \\
\multirow{-4}{*}{GPT-5.5}
 & \evshade \tool & \evshade 102 & \evshade \textbf{0} & \evshade 1 (0) & \evshade 0.36 & \evshade 0.00 & \evshade 71.0 \\
\midrule

 & BugLens          & 127 & 14 & 0      & 0.21 & 0.35 & 76.5 \\
\multirow{-2}{*}{Claude Sonnet 4.6}
 & \evshade \tool & \evshade 98 & \evshade \textbf{0} & \evshade 14 (1) & \evshade 0.39 & \evshade 0.00 & \evshade 69.0 \\
\midrule

 & BugLens          & 132 & 17 & 2 (2)  & 0.18 & 0.43 & 77.5 \\
\multirow{-2}{*}{GLM-5.1}
 & \evshade \tool & \evshade 94 & \evshade \textbf{0} & \evshade 23 (2) & \evshade 0.41 & \evshade 0.00 & \evshade 67.0 \\
\midrule

 & BugLens          & 28  & 4 & 0       & 0.83 & 0.10 & 32.0 \\
\multirow{-2}{*}{GPT-5.4 mini}
 & \evshade \tool & \evshade 78 & \evshade \textbf{0} & \evshade 47 (3) & \evshade 0.51 & \evshade 0.00 & \evshade 59.0 \\

\bottomrule
\end{tabular}
\end{table}

\begin{figure}[t]
\centering
\begin{tikzpicture}
\begin{axis}[
  width=0.80\linewidth, height=0.50\linewidth,
  xlabel={False alarms filtered out (of 160)},
  ylabel={Confirmed bugs kept (of 40)},
  xmin=0, xmax=160, ymin=0, ymax=40,
  enlargelimits=false,
  xtick={0,40,80,120,160}, ytick={0,10,20,30,40},
  axis lines*=left, axis line style={gray!60},
  grid=major, grid style={gray!15},
  tick label style={font=\scriptsize}, label style={font=\footnotesize},
  legend style={at={(0.02,0.03)}, anchor=south west, font=\scriptsize,
    draw=none, fill=none, cells={anchor=west}},
  clip=false,
]
\addplot[gray!60, thin, forget plot] coordinates {(0,40) (160,40)};
\node[font=\tiny, text=gray!80, anchor=south east, inner sep=1.5pt] at (axis cs:160,40)
  {all 40 confirmed bugs kept};
\addplot[gray!70, dotted, thin, forget plot] coordinates {(0,40) (160,0)};
\addplot[only marks, mark=o, mark size=2.2pt, gray!80, forget plot]
  coordinates {(0,40) (160,0)};
\node[font=\tiny, text=gray!80, anchor=south, rotate=-29] at (axis cs:50,28.3)
  {random filtering};
\node[font=\tiny, text=gray!80, anchor=north west] at (axis cs:1.5,39.2) {keep all};
\node[font=\tiny, text=gray!80, anchor=south east] at (axis cs:158.5,0.6) {filter out all};
\addplot[only marks, mark=*, mark size=2.2pt, black]
  coordinates {(102,40) (98,40) (94,40) (78,40)};
\addlegendentry{\tool (four LLMs)}
\addplot[only marks, mark=square*, mark size=2.0pt, figblue]
  coordinates {(127,26) (132,23) (28,36)};
\addlegendentry{BugLens (four LLMs)}
\addplot[only marks, mark=triangle*, mark size=2.8pt, figochre]
  coordinates {(124,22) (123,29)};
\addlegendentry{Codex (GPT-5.5)}
\draw[gray!60, thin] (axis cs:78,39.2)  -- (axis cs:74,37.4);
\draw[gray!60, thin] (axis cs:94,39.2)  -- (axis cs:90.5,37.4);
\draw[gray!60, thin] (axis cs:98,39.2)  -- (axis cs:98,35.0);
\draw[gray!60, thin] (axis cs:102,39.2) -- (axis cs:105.5,37.4);
\node[font=\tiny, anchor=north east] at (axis cs:74.5,37.2)  {GPT-5.4 mini};
\node[font=\tiny, anchor=north east] at (axis cs:91,37.2)    {GLM-5.1};
\node[font=\tiny, anchor=north]      at (axis cs:98,34.8)    {Sonnet 4.6};
\node[font=\tiny, anchor=north west] at (axis cs:105,37.2)   {GPT-5.5};
\node[font=\tiny, anchor=north] at (axis cs:28,35)    {GPT-5.4 mini};
\node[font=\tiny, anchor=east]  at (axis cs:125,26)   {GPT-5.5, Sonnet 4.6};
\node[font=\tiny, anchor=west]  at (axis cs:134,23)   {GLM-5.1};
\node[font=\tiny, anchor=north] at (axis cs:124,21.2) {Codex};
\node[font=\tiny, anchor=west]  at (axis cs:125.5,29) {Codex (FN-aware)};
\end{axis}
\end{tikzpicture}
\caption{Safety and utility of each system in Table~\ref{tab:llm-filtering}.
Each point is one system with one LLM; the upper right is best.}
\label{fig:tradeoff}
\end{figure}

\subsubsection{Baselines}
\label{sec:eval-baselines}
We evaluate all baselines on the same 200 warnings and ground
truth as \tool. Below, we describe their configurations and
the adjustments made for a consistent comparison.

\textbf{BugLens.}
We evaluate BugLens with the same four LLMs used by \tool.
We retain its original scaffold and repository-level code retrieval.
BugLens also considers security impact when assessing warnings,
whereas our ground truth concerns whether the reported error can occur.
We therefore disable its security-impact assessment so that both
systems are evaluated against the same criterion.

\textbf{Codex.}
We evaluate Codex with GPT-5.5 at \texttt{xhigh} reasoning.
For each warning, Codex receives the detector report, including
the reported location, call chain, and available trace information.
It also has read-only access to the complete kernel source tree
for the analyzed version and can independently investigate
the warning using shell commands and code search, following
data and control flow across functions and files.
Codex is asked whether the reported error can occur, allowing
adversarial user inputs but assuming an uncompromised kernel.
Its decisions must be grounded in the supplied report and
inspected source code; external sources, ground-truth labels,
and previous evaluation results are excluded.
Codex may filter out a warning, keep it, or abstain when
the available evidence is insufficient.

We evaluate two configurations with the same model, warning
information, and source access.
The standard configuration already permits abstention when
evidence is insufficient.
The \emph{FN-aware} variant additionally instructs Codex to
prioritize avoiding false negatives, exercise greater caution when filtering out warnings, and keep the warning or
abstain when uncertain.

\subsubsection{Results}
\label{sec:eval-tradeoff-results}

Figure~\ref{fig:tradeoff} compares how many false alarms and how many
confirmed bugs each system filters out.

With GPT-5.5, \tool filters out 102 of the 160 false alarms
and none of the 40 confirmed bugs. BugLens filters out 127
false alarms but also 14 confirmed bugs, while Codex filters out
124 false alarms but also 18 confirmed bugs. Compared with
these baselines, \tool keeps 22--25 more false alarms but filters
out no confirmed bug.

Explicitly instructing Codex to avoid false negatives reduces FN from
18 to 11, while TN decreases only slightly, from 124 to 123.
Thus, conservative instructions help, but Codex still filters out
confirmed bugs.

In terms of overall accuracy, \tool achieves 71.0\% with GPT-5.5,
compared with 76.5\% for BugLens, 73.0\% for Codex, and 76.0\%
for FN-aware Codex.
Compared with BugLens and FN-aware Codex, \tool sacrifices
5.0--5.5 percentage points in accuracy while avoiding the
11--14 confirmed bugs these baselines incorrectly filter out.

With Sonnet 4.6 and GLM-5.1, BugLens filters out more false alarms
than \tool but also 14 and 17 confirmed bugs.
With GPT-5.4 mini, BugLens filters out only 28 false alarms,
fewer than \tool's 78, and still four confirmed bugs.
Here, \tool also achieves higher overall accuracy
(59.0\% versus 32.0\%).
Across all four LLMs, \tool keeps every confirmed bug,
whereas BugLens filters out confirmed bugs with every model.

\begin{takeaway}{RQ2}
Compared with recent LLM-decided approaches, \tool leaves 13--24 percentage
points more of the false alarms for inspection with the three stronger LLMs,
and filters out no confirmed bug with any LLM, against 4--18 of the 40 for the
baselines.
\end{takeaway}

\subsection{RQ3: Design Choices}
\label{sec:design-variants}

RQ3 examines three design choices that limit the LLM's control over the
outcome: warning-relevant inputs take abstract values rather than values the
LLM picks, an analysis result counts only if the harness passes validation,
and \eva's result, not the LLM, decides whether a warning is filtered out.
Table~\ref{tab:design-choices} compares four configurations, all with
GPT-5.5. (D) is the full system. (A) drops the prompt instruction that
requires \texttt{abstract\_value}, so the LLM chooses the input values, and
runs without validation and repair. (B) keeps that instruction but still
runs without validation and repair. (C) takes the harnesses and \eva outputs
of (D) but lets the LLM make the final decision instead.

\begin{table}[t]
\centering
\caption{Design choices and outcomes with GPT-5.5; unresolved column as in
Table~\ref{tab:llm-filtering}.}
\label{tab:design-choices}
\small
\setlength{\tabcolsep}{3pt}
\begin{tabular}{l|ccc|rr|r|rr}
\toprule
\multirow{2}{*}{\textbf{Variant}} &
\multicolumn{3}{c|}{\textbf{\textit{Design}}} &
\multicolumn{2}{c|}{\textbf{\textit{Filtered out}}} &
\multirow{2}{*}{\textbf{Unres.}} &
\multicolumn{2}{c}{\textbf{\textit{Rates}}} \\
& \textbf{Input}
& \textbf{Validation}
& \textbf{Decision}
& \textbf{TN}
& \textbf{FN}
&
& \textbf{FPR}
& \textbf{FNR} \\
\midrule

(A) LLM-decided input
& LLM
& --
& Backend
& 116 & 9 & 6 (0) & 0.28 & 0.23 \\

(B) Abstract input
& Abstract
& --
& Backend
& 99 & 1 & 5 (0) & 0.38 & 0.03 \\

(C) LLM final decision
& Abstract
& \checkmark
& LLM
& 98 & 4 & 0 & 0.39 & 0.10 \\

\midrule
(D) Full \tool
& Abstract
& \checkmark
& Backend
& 102 & \textbf{0} & 1 (0) & 0.36 & 0.00 \\

\bottomrule
\end{tabular}
\end{table}

\textbf{Input initialization.}
In (A), the LLM may initialize a warning-relevant value with a concrete
value, interval, or backend-specific expression of its choice.
Eight of the nine FN in (A) arise because the generated harness restricts
such a value too narrowly, allowing the analyzer to rule out the reported
error only under that restriction.
The same restrictions let the analyzer rule out more false alarms as well:
(A) filters out 116 of the 160, more than any other configuration.

Configuration (B) instead initializes these values using
\texttt{abstract\_value}, preserving the full range of their C types.
(B) keeps eight of the nine confirmed bugs that (A) filtered out; the ninth
remains filtered out. Overall, FN decreases from nine to one and TN from
116 to 99.

\textbf{Harness validation.}
Configuration (D) enables the validation checks of
\S\ref{sec:design-validation} and the repair loop following
a validation failure.
The checks detect the harness error responsible for the only FN in (B).
With validation and its repair feedback enabled, (D) keeps all 40 confirmed
bugs. Unresolved runs decrease from five to one, and TN increases from
99 to 102.
Applying the checks and repair to the harnesses generated in (B) also keeps
all 40 confirmed bugs. The zero FN of (D) therefore does not depend on
generating new harnesses.

\textbf{Final decision.}
The LLM-decided baselines in RQ2 filter out more false alarms,
suggesting that an LLM might filter out more if it also saw
the analyzer's results.
To test this, we keep the harnesses and \eva outputs of the
full GPT-5.5 run but ask GPT-5.5 to make the final decision.
The model receives the warning, generated harness, analyzer results,
and access to the kernel source, and returns
\texttt{not\_a\_bug}, \texttt{still\_a\_bug}, or \texttt{uncertain}.

Compared with (D), the LLM filters out four fewer false alarms
(98 versus 102) while also filtering out four confirmed bugs.
In three of these four cases, the LLM reasons incorrectly about C integer
semantics; in the fourth, it discounts an \eva alarm after judging the
reported operation itself safe.

\begin{takeaway}{RQ3}
Letting the LLM choose input values filters out more false alarms, but the
ranges it chooses are too narrow and also hide confirmed bugs from \eva.
Validation catches the harness errors that remain and lets the LLM repair
them. Letting the LLM make the final decision filters out no more false
alarms than \eva, and it filters out confirmed bugs that \eva had flagged.
\end{takeaway}


\subsection{Case Studies}
\label{sec:case-studies}

To complement the aggregate results, we present two cases
illustrating why \tool leaves bug decisions to program analysis.
In the first, a branch visible in the source is unreachable
under the analyzed build configuration.
In the second, an LLM identifies the information needed for
an array-bounds check, but leaves the actual decision to \eva.

\textbf{Code-as-seen is not code-as-run.}
In one warning, the reported operation is guarded by a compound condition
roughly 100 lines earlier in a roughly 400-line function. Reasoning from the
source appears to require understanding several conditions involving
framework versions and hardware support. One of these conditions, however,
depends on the build configuration:

\begin{Verbatim}[commandchars=\\\{\},numbers=left,firstnumber=1,numbersep=5pt,xleftmargin=1.6em,fontsize=\footnotesize]
\PYG{c+cp}{\PYGZsh{}if IS\PYGZus{}ENABLED(CONFIG\PYGZus{}USE\PYGZus{}Q6\PYGZus{}32CH\PYGZus{}SUPPORT)}
\PYG{k+kt}{bool}\PYG{+w}{ }\PYG{n+nf}{q6core\PYGZus{}use\PYGZus{}Q6\PYGZus{}32ch\PYGZus{}support}\PYG{p}{()}\PYG{+w}{ }\PYG{p}{\PYGZob{}}\PYG{+w}{ }\PYG{k}{return}\PYG{+w}{ }\PYG{n+nb}{true}\PYG{p}{;}\PYG{+w}{ }\PYG{p}{\PYGZcb{}}
\PYG{c+cp}{\PYGZsh{}else}
\PYG{k+kt}{bool}\PYG{+w}{ }\PYG{n+nf}{q6core\PYGZus{}use\PYGZus{}Q6\PYGZus{}32ch\PYGZus{}support}\PYG{p}{()}\PYG{+w}{ }\PYG{p}{\PYGZob{}}\PYG{+w}{ }\PYG{k}{return}\PYG{+w}{ }\PYG{n+nb}{false}\PYG{p}{;}\PYG{+w}{ }\PYG{p}{\PYGZcb{}}\PYG{esc}{\phantomsection\label{ln:case1false}}
\PYG{c+cp}{\PYGZsh{}endif}
\end{Verbatim}

In the analyzed build,
\texttt{CONFIG\_USE\_Q6\_32CH\_SUPPORT} is disabled, so the helper returns
\texttt{false} on Line~\ref{ln:case1false} and the reported branch cannot
execute. An LLM reasoning from
the visible source can instead treat the branch as feasible unless it also
finds and resolves this configuration choice.



\tool instead compiles the harness under the analyzed kernel
configuration, allowing \eva to establish that the reported
branch is unreachable.
Providing the configuration to an LLM or preprocessing the
source could also resolve this particular case, but neither
guarantees correct reasoning about program reachability
in general.

\textbf{Leaving the capacity check to Eva.}
SUTURE reports a potential out-of-bounds write in
\texttt{msm\_compr\_dec\_params\_put}
(warning \texttt{msm-sound:0066}).
The function reads a user-supplied count and copies that many
pairs of values into two arrays.
In the code below, we mark the shared array capacity as \(C\)
and the count limit as \(L\).
Declarations are compacted and unrelated guards omitted.

\begingroup
\newcommand{\caseMacro}[2]{%
  {\setlength{\fboxsep}{1pt}%
   \colorbox{blue!8}{\strut\texttt{#1}}%
   \textsuperscript{\textsf{\scriptsize #2}}}}
\begin{Verbatim}[commandchars=\\\{\},numbers=left,firstnumber=1,numbersep=5pt,xleftmargin=1.6em,fontsize=\footnotesize]
\PYG{k}{struct}\PYG{+w}{ }\PYG{n+nc}{snd\PYGZus{}dec\PYGZus{}ddp}\PYG{+w}{ }\PYG{p}{\PYGZob{}}\PYG{+w}{ }\PYG{n}{\PYGZus{}\PYGZus{}u32}\PYG{+w}{ }\PYG{n}{params\PYGZus{}length}\PYG{p}{;}\PYG{esc}{\phantomsection\label{ln:case2type}}
\PYG{+w}{    }\PYG{n}{\PYGZus{}\PYGZus{}u32}\PYG{+w}{ }\PYG{n}{params\PYGZus{}id}\PYG{p}{[}\PYG{esc}{\caseMacro{SND\_DEC\_DDP\_MAX\_PARAMS}{C}}\PYG{p}{],}\PYG{esc}{\phantomsection\label{ln:case2capacity}}
\PYG{+w}{          }\PYG{n}{params\PYGZus{}value}\PYG{p}{[}\PYG{esc}{\caseMacro{SND\_DEC\_DDP\_MAX\_PARAMS}{C}}\PYG{p}{];}\PYG{+w}{ }\PYG{p}{\PYGZcb{};}\PYG{esc}{\phantomsection\label{ln:case2capacityvalue}}
\PYG{n}{ddp}\PYG{o}{\PYGZhy{}\PYGZgt{}}\PYG{n}{params\PYGZus{}length}\PYG{+w}{ }\PYG{o}{=}\PYG{+w}{ }\PYG{o}{*}\PYG{n}{values}\PYG{o}{++}\PYG{p}{;}\PYG{esc}{\phantomsection\label{ln:case2input}}
\PYG{k}{if}\PYG{+w}{ }\PYG{p}{(}\PYG{n}{ddp}\PYG{o}{\PYGZhy{}\PYGZgt{}}\PYG{n}{params\PYGZus{}length}\PYG{+w}{ }\PYG{o}{\PYGZgt{}}\PYG{+w}{ }\PYG{esc}{\caseMacro{DDP\_DEC\_MAX\_NUM\_PARAM}{L}}\PYG{p}{)}\PYG{+w}{ }\PYG{k}{goto}\PYG{+w}{ }\PYG{n}{end}\PYG{p}{;}\PYG{esc}{\phantomsection\label{ln:case2guard}}
\PYG{k}{for}\PYG{+w}{ }\PYG{p}{(}\PYG{n}{cnt}\PYG{+w}{ }\PYG{o}{=}\PYG{+w}{ }\PYG{l+m+mi}{0}\PYG{p}{;}\PYG{+w}{ }\PYG{n}{cnt}\PYG{+w}{ }\PYG{o}{\PYGZlt{}}\PYG{+w}{ }\PYG{n}{ddp}\PYG{o}{\PYGZhy{}\PYGZgt{}}\PYG{n}{params\PYGZus{}length}\PYG{p}{;}\PYG{+w}{ }\PYG{n}{cnt}\PYG{o}{++}\PYG{p}{)}\PYG{+w}{ }\PYG{p}{\PYGZob{}}\PYG{esc}{\phantomsection\label{ln:case2loop}}
\PYG{+w}{    }\PYG{n}{ddp}\PYG{o}{\PYGZhy{}\PYGZgt{}}\PYG{n}{params\PYGZus{}id}\PYG{p}{[}\PYG{n}{cnt}\PYG{p}{]}\PYG{+w}{ }\PYG{o}{=}\PYG{+w}{ }\PYG{o}{*}\PYG{n}{values}\PYG{o}{++}\PYG{p}{;}\PYG{esc}{\phantomsection\label{ln:case2write}}
\PYG{+w}{    }\PYG{n}{ddp}\PYG{o}{\PYGZhy{}\PYGZgt{}}\PYG{n}{params\PYGZus{}value}\PYG{p}{[}\PYG{n}{cnt}\PYG{p}{]}\PYG{+w}{ }\PYG{o}{=}\PYG{+w}{ }\PYG{o}{*}\PYG{n}{values}\PYG{o}{++}\PYG{p}{;}\PYG{esc}{\phantomsection\label{ln:case2writevalue}}
\PYG{p}{\PYGZcb{}}
\end{Verbatim}

\endgroup

Both macros expand to 18, so \(L=C=18\).
The guard on Line~\ref{ln:case2guard} ensures that the loop
accesses only indices 0--17, within both arrays.
The function also reads at most \(1+2\times18=37\) entries
from the 128-entry control input (\texttt{*values}).



Through BugLens, GPT-5.4 mini retrieved the structure definition
and recognized that the count was bounded by \(L\).
It also identified the need to compare \(L\) with the array
capacity \(C\), but did not complete this comparison.
Although BugLens could retrieve both macro definitions,
the model made no further queries and returned
\texttt{still\_a\_bug}.
A stronger LLM may  complete this comparison, but \tool
does not rely on the model to perform such checks correctly.

With \tool, the same model constructed a harness that connects
the control, component, and device to the driver's private data
and initializes the required runtime objects.
The repaired harness calls
\texttt{msm\_compr\_dec\_params\_put} directly.
Its selector, codec, passthrough mode, and all 128 input slots
use \texttt{abstract\_value}; validation checks confirmed their
full numeric type ranges at entry.

Compiling the harness with the driver provides \eva with the
actual macro values and array layouts.
The analysis reached both the count assignment and the loop (2/2)
and completed with zero alarms.
\tool therefore correctly filtered out the warning.

\section{Discussion \& Limitations}
\label{sec:discussion}

\textbf{From LLM Judgments to Checkable Evidence.}
\tool uses the LLM to reconstruct analysis context, but leaves the final
warning decision to the analyzer.
This choice is deliberately conservative: LLM-decided approaches often
filter out more false alarms, but they filter out confirmed bugs with
every evaluated model.
RQ3 further shows that handing the final decision back to the LLM after
analysis does not filter out more false alarms.
The LLM largely follows the analyzer's results, but its occasional
deviations filter out confirmed bugs.

More importantly, \tool turns assumptions implicit in LLM judgments
into \textit{checkable} code: validation can reject invented input
bounds and detect unconnected callbacks through gaps in the
reported value flow.
Our current three checks in \S\ref{sec:design-validation} target construction
errors we observed, but an explicit harness is always available for additional checks.
The harness, the validation results, and the analyzer output can be
inspected independently of the LLM; an LLM's no-bug explanation alone
provides no comparable evidence.


\textbf{Scope.}
We evaluate \tool on Android kernel drivers, which present a challenging
setting for static analysis.
Driver functions often depend on kernel framework callbacks,
complex program state, API behavior, and build configurations,
making them difficult to analyze in isolation.
These characteristics motivate \tool's use of LLMs to reconstruct
warning-specific analysis contexts.
Similar challenges arise in other context-dependent software,
such as libraries and framework-based applications, where the
approach may also be applicable.

\textbf{Limitations and Threats to Validity.}
\emph{Internal validity.}
We run each model configuration three times, which captures only a limited
estimate of LLM nondeterminism. Our validation checks target selected harness
properties; construction errors outside these checks could still cause confirmed
bugs to be filtered out.

\emph{External validity.}
Our evaluation focuses on Android kernel drivers and the warning classes
supported by our front end.
Results may differ for proprietary or less familiar code bases,
other bug classes, or other analysis backends.

\section{Related Work}

\textbf{Reliability of LLM reasoning for security.}
Recent studies show that LLMs remain unreliable as direct vulnerability
detectors: models can misclassify buggy and patched code, produce
setting-dependent or non-deterministic results, and give incorrect rationales for
security judgments~\cite{steenhoek_err_2025,ullah_llms_2024,
khare_understanding_2024,chen2025reasoning,fang_large_2024}. \tool uses the
model's help without trusting its judgment: a no-bug result requires backend analysis of an
accepted harness.

\textbf{LLM-based warning filtering.}
Recent work has shown that LLMs can be effective at filtering static-analysis
warnings~\cite{wen_automatically_2024,chen2024utilizing,
li_iris_2025,buglens,llift}. These systems improve over bare warning reports by
supplying the model with code context, slices, paths, or structured reasoning
scaffolds, and they report strong empirical results on warning classification.
Similar ideas have also appeared in industrial SAST workflows, where LLMs
classify findings as likely true or false positives~\cite{maring_using_2025,
du2026reducing-fp}.

\tool follows the same insight that LLMs are useful for recovering
the context relevant to a warning, but uses that context differently. Prior
systems use the recovered context to support the model's own decision. \tool
instead turns it into a warning-specific analysis harness and decides only
after backend analysis of that harness.

\textbf{Grounding LLMs with program analysis.}
\textsc{LLM4PFA}~\cite{du2025minimizing} checks source-to-sink
feasibility through path-constraint solving.
\textsc{LLMDFA} and \textsc{RepoAudit} combine LLM reasoning
with data-flow facts and solver-backed path validation
~\cite{DBLP:conf/nips/WangZSXX024,DBLP:journals/corr/abs-2501-18160}.
Concurrent work guides symbolic execution with LLM-generated
harnesses and confirms bugs through concrete replay
~\cite{shafiuzzaman2026guiding}.
These approaches seek a feasible execution to confirm a bug.
Filtering out a warning instead requires ruling out all executions
that trigger the reported error, which a generated harness might
inadvertently exclude.
\tool therefore keeps warning-relevant inputs unconstrained and
checks that the harness preserves the reported value flow before accepting a
no-bug result.


\textbf{LLM-based fuzz-driver generation.}
LLMs have been used to synthesize fuzzing harnesses and drivers for dynamic
bug finding~\cite{ossfuzzgen,lyu2024prompt,liu2025promefuzz,xu2025ckgfuzzer,
li2025scheduzz,jeong2023utopia}. \tool also asks an LLM to generate a harness,
but for a different purpose: the harness is a warning-specific artifact for
backend analysis, not a driver for finding new crashes. Its validation is
therefore tied to the reported warning rather than to compilation, coverage, or
fuzzing yield.

\section{Conclusion}
\label{sec:conclusion}

Program analysis exists to analyze programs. This is almost tautological:
judgments about program behavior should come from analysis, not from guesses
about the code. Yet the current enthusiasm for LLMs makes this boundary easy to
blur. This paper asks where LLMs should sit in bug analysis so that the tautology
still holds.

Our answer is to use the LLM to write down, as a program, the context that the
backend could not otherwise analyze.
Making this context an explicit program also turns the remaining soundness
question into something concrete and open to independent checks. \tool
demonstrates that this approach is practical on real Android Linux kernel
drivers: LLMs can carry much of the context-construction burden while program
analysis remains responsible for deciding program behavior. \tool requires evidence before it filters out a warning, and keeps the warning
when it cannot produce that evidence.

More broadly, using an LLM for more of an analysis pipeline is not necessarily
better. When established program analysis provides a stronger basis for a
program-behavior decision, the LLM can instead be used to make that analysis
possible.



\bibliographystyle{ACM-Reference-Format}
\bibliography{main}

\end{document}